\documentclass[a4paper]{article}

\usepackage{xcolor}
\definecolor{solarizedYellow}{HTML}{B58900}
\definecolor{solarizedRed}{HTML}{DC322F}
\definecolor{solarizedMagenta}{HTML}{D33682}
\definecolor{solarizedViolet}{HTML}{6C71C4}
\definecolor{solarizedBlue}{HTML}{268BD2}
\definecolor{solarizedCyan}{HTML}{2AA198}
\definecolor{solarizedGreen}{HTML}{859900}

\usepackage{hyperref}

\usepackage[utf8]{inputenc}

\usepackage{float}
\usepackage{booktabs}
\usepackage{adjustbox}
\usepackage{hyperref}
\usepackage{amsmath,amssymb}
\usepackage[natbib,cref,theorems]{nikis}

\newcommand*{\functionname}[1]{{{\renewcommand{\rmdefault}{ptm}\fontfamily{ppl}\selectfont\textrm{\textup{#1}}}}} \newcommand*{\fnLocate}{\functionname{locate}}
\newcommand*{\fnCount}{\functionname{count}}
\newcommand*{\fnExists}{\functionname{exists}}

\newcommand*{\typeL}{\texttt{L}}
\newcommand*{\typeS}{\texttt{S}}
\newcommand*{\typeHoshi}{\ensuremath{\texttt{S}^*}}

\newcommand*{\instancename}[1]{\ensuremath{\mathsf{#1}}} 

\newcommand*{\BWT}{\instancename{BWT}}
\newcommand*{\Last}{\instancename{Last}}

\definecolor{teigiIro}{HTML}{5700B5}
\newcommand*{\teigi}[1]{{\color{teigiIro}\emph{#1}}} 

\newcommand{\select}[1][]{\operatorname{select}_{#1}}
\newcommand{\rank}[1][]{\operatorname{rank}_{#1}}

\newcommand*{\bv}[1]{\ensuremath{B_{\mathup{#1}}}}

\newcommand*{\Takai}[2]{\ensuremath{{#1}^{(#2)}}}

\newcommand*{\BunpouT}{\ensuremath{\mathcal{G}_T}}
\newcommand*{\BunpouP}{\ensuremath{\mathcal{G}_P}}
\newcommand*{\Symbols}{\ensuremath{\mathcal{S}}}
\newcommand*{\RLFM}[1]{\ensuremath{\instancename{RLFM}^{(#1)}}}

\usepackage{ifthen}
\usepackage{forest}
\usepackage{tikz}
\usetikzlibrary{matrix,fit,positioning}
	\makeatletter
	\tikzset{store number of columns in/.style={execute at end matrix={
		\xdef#1{\the\pgf@matrix@numberofcolumns}}},
		store number of rows in/.style={execute at end matrix={
	\xdef#1{\the\pgfmatrixcurrentrow}}}}
	\makeatother

\newcommand*{\LMSbracket}[2]{\begin{scope}[transform canvas={yshift=-0.4em,xshift=-0.1em}]
		\draw  ([yshift=0.4em]text-3-#1.south) -- (text-3-#1.south) -- ([xshift=-0.1em]text-3-#2.south) --  ([xshift=-0.1em,yshift=0.4em]text-3-#2.south)  {};
	    \end{scope}
	}

\tikzset{global scale/.style={scale=#1,
    every node/.style={scale=#1}
  }
}

\tikzstyle{array} = [matrix of nodes
    , nodes in empty cells
	, column sep=0.5\pgflinewidth
	, row sep=0.5mm
	, store number of columns in=\mymatcols
	, row 1/.style={nodes={font=\ttfamily,scale=0.9, fill=none, minimum size=0mm, text height=0.5em}}
	, row 2/.style={nodes={font=\tiny,fill=none, minimum size=0mm, text height=0.33em}}
	]
\tikzstyle{Pattern} = [array, store number of columns in=\mypatterncols]
\tikzstyle{Text} = [array, store number of columns in=\mytextcols]
\tikzstyle{Block} = [inner sep=0, inner xsep=5, draw=red,rounded corners=4pt]

\usepackage{multirow} 

\begin{document}

\title{\Large FM-Indexing Grammars Induced by Suffix Sorting for Long Patterns}
\author{Anonymous}
  \author{Jin Jie Deng\thanks{Department of Computer Science, National Tsing Hua University, Hsinchu 30013, Taiwan, jinjiedeng.jjd@gmail.com}
  \and 
  Wing-Kai Hon\thanks{Department of Computer Science, National Tsing Hua University, Hsinchu 30013, Taiwan, wkhon@cs.nthu.edu.tw}
  \and 
  Dominik K\"{o}ppl\thanks{M\&D Data Science Center, Tokyo Medical and Dental University, Japan, koeppl.dsc@tmd.ac.jp}
  \and 
  Kunihiko Sadakane\thanks{The University of Tokyo, Japan, sada@mist.i.u-tokyo.ac.jp}
}

\date{}

\maketitle

\begin{abstract} \small\baselineskip=9ptThe run-length compressed Burrows-Wheeler transform (RLBWT) used in conjunction with the backward search introduced in the FM index is the centerpiece of most compressed indexes working on highly-repetitive data sets like biological sequences.
Compared to grammar indexes, the size of the RLBWT is often much bigger, but queries like counting the occurrences of long patterns can be done much faster than on any existing grammar index so far.
In this paper, we combine the virtues of a grammar with the RLBWT by building the RLBWT on top of a special grammar based on induced suffix sorting.
Our experiments reveal that our hybrid approach outperforms the classic RLBWT with respect to the index sizes, and with respect to query times on biological data sets for sufficiently long patterns.
\end{abstract}

\section{Introduction}
A text index built on a string $T$ of length~$n$ is a data structure that can answer the following queries, for a given pattern~$P$ of length~$m$:
\begin{description}
  \crampedItems
  \item[\fnExists(P)]: does the pattern~$P$ occur in $T$?
  \item[\fnCount(P)]: how often does the pattern~$P$ occur in $T$?
  \item[\fnLocate(P)]: where does the pattern~$P$ occur in $T$?
\end{description}
The answers are a boolean, a number, and a list of starting positions in the text, respectively.
$\fnLocate(P)$ is the most powerful query because the cardinality of its returned set is the return value of $\fnCount(P)$, 
whereas $\fnCount(P) > 0$ is a boolean statement equivalent to $\fnExists(P)$.

One prominent example of such a text index is the FM-index~\cite{ferragina00fmindex}.
It consists of a wavelet tree~\cite{grossi03wavelet} built upon the BWT~\cite{burrows94bwt} of the text, and
can answer $\fnCount(P)$ in time linear to the length of $P$ multiplied by the operational cost of the wavelet tree, 
which can be logarithmic in the alphabet size and up to constant~\cite{belazzougui14index}.
Given the BWT consists of $r$ maximal character runs, this data structure can be represented by two additional bit vectors~\cite[Thm.~3]{makinen05rle} of length~$n$ in $r \lg \sigma + \oh{r \lg \sigma} + \Oh{n}$ bits of space.
This space can be further reduced with Huffman-shaped wavelet trees by exploiting the zeroth order empirical entropy on the string consisting of the different letters of the runs in the BWT\@.
For $\fnLocate$, the indexes based on the BWT are augmented by a sampling of the suffix array~\cite{manber93sa}, which needs $n \lg n$ bits in its plain form. 
In what follows, we do not address $\fnLocate$ since this augmentation can be done orthogonal to our proposed data structure, and is left as future work.

Although current approaches achieve \Oh{m} time for $\fnCount(P)$ with $|P| = m$, 
it involves $\Oh{m}$ queries to the underlying wavelet tree data structure, which is performed in a constant number of random accesses.
Unfortunately, these random accesses make the FM-index rather slow in practice.
The BWT built on a grammar compressed string allows us to match non-terminals in one backward search step, hence
allowing us to \emph{jump over multiple characters} in one step.
Consequently, we spend less time on the cache-unfriendly wavelet tree, 
but more time on extracting the grammar symbols stored in cache-friendly arrays.
Our experiments reveal that this extra work pays off for the reduced usage of the wavelet tree regarding the time performance.
Regarding the space, the grammar captures the compressibility far better than the run-length compression of the BWT built on the plain text.
Here, we leverage certain properties of the GCIS (grammar compression by induced suffix sorting) grammar~\cite{nunes18grammar}, 
which have been discovered by~\citet{akagi21grammar} and \citet{diaz21lms} for determining non-terminals of the text matching portions of the pattern.

\paragraph{Our Contribution}
To sum up, our contribution is that
combining the BWT with a specific choice of grammar-based compression method achieves potentially better compression than the plain RLBWT,
and at the same time reducing the memory accesses for count queries (heuristically).
This comes at the expense of additional computation for building the grammar of the text during the construction and of the pattern during a query.

\subsection*{Related Work}
A lot of research effort has been invested in analyzing and improving \fnCount{} of the BWT (e.g., \cite{makinen05rle} and the references therein) and the sampling of the suffix array (e.g.,~\cite{gagie18bwt} and the references therein).
Another line of research are grammar indexes, which usually enhance a grammar for \fnLocate{} queries.
Although computing the smallest grammar is NP-complete~\cite{charikar05grammar}, there are grammars with a size of \Oh{r \log(n/r)}~\cite{gagie18bwt},
and some grammars are empirically much smaller than the RLBWT in practice.
However, most indexes have a quadratic dependency on the pattern length for \fnLocate{}~\cite{claude12grammarindex,claude21grammarcompressed}, 
and are unable to give improved query times independent of the number of occurrences of the pattern, when considering only \fnCount{}.
A novel exception is the grammar index of \citet{christiansen21optimaltime}, which achieves \Oh{m + \log^{2+\epsilon} n} time for \fnCount{}
with a space of $\Oh{\gamma \log(n/\gamma)}$, for $\gamma$ being the size of the smallest string attractor~\cite{kempa18stringattractors} of the input text. 
However, this approach seems to be rather impractical, and up to now nobody has considered implementing it.
Related to our work is the grammar indexes of \citet{akagi21grammar} and \citet{diaz21lms}, which are also based on the GCIS grammar, where the latter is based on results of \citet{christiansen21optimaltime}.
They also use similar techniques for extracting non-terminals from the pattern grammar,
for which they can be sure of that these appear in the text grammar.
However, they need to call \fnLocate{} for computing \fnCount{}, and thus their time complexity is dependent on the number of occurrences of a pattern.

We are not aware of a combination of the BWT with grammar techniques, except for construction.
Here, \citet{karkkainen12bwt} studied the construction of the BWT upon a grammar-compressed input.
They applied a grammar compression merging frequent bigrams similar to Re-Pair~\cite{larsson99repair}, 
and empirically could improve the computation of the BWT as well as the reconstruction of the text from the BWT\@.
With a similar target, \citet{diazdominguez21grammar,diaz21grammar} computed the extended BWT~\cite{mantaci07ebwt}, a BWT variant for multiple texts, from the GCIS grammar.

\section{Preliminaries}

With $\lg$ we denote the logarithm to base two (i.e., $\lg = \log_2$).
Our computational model is the word RAM with machine word size~$\Om{\lg n}$,
where $n$ denotes the length of a given input string~$T[1..n]$, which we call \teigi{the text},
whose characters are drawn from an integer alphabet~$\Sigma = \{1,\ldots,\sigma\}$ of size $\sigma = n^{\Oh{1}}$. 
We call the elements of $\Sigma$ \teigi{characters}.
A \teigi{character run} is a maximal substring consisting of repetition of the same character.
For a string $S \in \Sigma^*$, we denote with $S[i..]$ its $i$-th suffix, and
with $|S|$ its length.
Given $X,Y,Z \in \Sigma^*$ with $S = XYZ$, 
then $X$, $Y$ and $Z$ are called a \teigi{prefix}, \teigi{substring} and \teigi{suffix} of $S$, respectively.
We say that a prefix~$X$ (resp.\ suffix~$Z$) is \teigi{proper} if $X \not= S$ (resp.\ $Z \not= S$).
The order $<$ on the alphabet~$\Sigma$ induces a lexicographic order on $\Sigma^*$, which we denote by $\prec$.

Given a character $\texttt{c} \in \Sigma$, and an integer~$j$, 
the \teigi{rank} query $T.\rank[\texttt{c}](j)$ counts the occurrences of \texttt{c} in $T[1..j]$, and 
the \teigi{select} query $T.\select[\texttt{c}](j)$ gives the position of the $j$-th \texttt{c} in $T$.
We stipulate that $\rank[\texttt{c}](0) = \select[\texttt{c}](0) = 0$.
If the alphabet is binary, i.e., when $T$ is a bit vector,
there are data structures~\cite{jacobson89rank,clark96select} that use \oh{|T|} extra bits of space, and
can compute $\rank{}$ and $\select{}$ in constant time, respectively.
Each of those data structures can be constructed in time linear in $|T|$. 
We say that a bit vector has a \emph{rank-support} and a \emph{select-support} if it is endowed 
by data structures providing constant time access to $\rank$ and $\select$, respectively.

\subsection{Burrows-Wheeler Transform}
The \teigi{BWT} of~$T$ is a permutation of the characters of $\tilde{T} := T\texttt{\$}$, 
where we appended an artificial character~$\texttt{\$}$ smaller than all characters appearing in $T$.
This BWT, denoted by \BWT{}, is defined such that $\BWT[i]$ is the preceding character of $\tilde{T}$'s $i$-th lexicographically smallest suffix, 
or $\tilde{T}[|\tilde{T}|] = \texttt{\$}$ in case that this suffix is $\tilde{T}$ itself.
Given a pattern $P[1..m]$, the \teigi{range} of $P[i..m]$ in $\BWT$ is an interval~$[\ell_i..r_i]$ such that 
$\tilde{T}[j..]$ has $P[i..m]$ as a prefix if and only if $\tilde{T}[j..]$ is the $k$-th lexicographically smallest suffix with $k \in [\ell_i..r_i]$.
The range~$[\ell_i,r_i]$ of~$P[i..m]$ can be computed from~$P[i+1..m]$ by a backward search step on \BWT{} with an array~$C[1..\sigma]$,
where $C[c]$ is the number of occurrences of those characters in~$\BWT$ that are smaller than $c$, for $c \in [1..\sigma]$.
Given the range of $P[i+1..m]$ is $[\ell_{i+1}..r_{i+1}]$, $\ell_i$ and $r_i$ are determined by
 $\ell_i = C[P[i]] + \BWT.\rank[{P[i]}](\ell_{i+1} + 1)$
 and 
 $r_i = C[P[i]] + \BWT.\rank[{P[i]}](r_{i+1})$,
 with $\ell_m = C[P[m]]+1$ and $r_m = C[P[m]+1]$.
We focus on ranges since the length of the range of $P$ is $\fnCount(P)$.

\paragraph{Example for BWT ranges}
Given the text from \cref{tabBWT}, and a pattern $P[1..6] = \texttt{cabaca}$, then
the range of $P[6] = \texttt{a}$ is $[2..6]$, and
the range of $P[5..6] = \texttt{ca}$ is $[12..13]$ since the \texttt{c}'s contained in the previous range $\BWT[2..6]$ 
are the second and third \texttt{c} in \BWT{}, which are in $F$ at positions $12$ and $13$,
where $F[i] = \mathop{\text{argmin}}_{c \in \Sigma} \{ C[c] : i \le C[c] \}$ is the $i$-th lexicographically smallest character in \BWT{}.

\begin{table}
  \centering
  \caption{BWT and array~$F$ of $T := \texttt{bacabacaacbcbc\$}$.}
  \label{tabBWT}
  \setlength{\tabcolsep}{3pt}
  \begin{tabular}{r*{15}{c}}
    \toprule
    $i$ & 1&2&3&4&5&6&7&8&9&10&11&12&13&14&15\\
    \midrule
    $F[i]$ & \texttt{\$} & \texttt{a}&\texttt{a}&\texttt{a}&\texttt{a}&\texttt{a}&\texttt{b}&\texttt{b}&\texttt{b}&\texttt{b}&\texttt{c}&\texttt{c}&\texttt{c}&\texttt{c}&\texttt{c}\\
    $\BWT[i]$ 
	& \texttt{c}&\texttt{c}&\texttt{c}&\texttt{b}&\texttt{b}&\texttt{a}&\texttt{a}& \texttt{\$} &\texttt{c}&\texttt{c}&\texttt{b}&\texttt{a}&\texttt{a}&\texttt{b}&\texttt{a}\\
    \bottomrule
  \end{tabular}
\end{table}

\subsection{Grammars}
An \teigi{admissible} (context-free) grammar~\cite{kieffer00code} built upon a string~$T \in \Sigma^*$ 
is a tuple $\BunpouT := (\Gamma, \pi, X_T)$ 
with $\Gamma$ being the set of non-terminals, 
a function $\pi : \Gamma \rightarrow (\Sigma \cup \Gamma)^+$ that applies (production) rules,
and a start symbol $X_T$ such that the iterative application of $\pi$ on $X_T$ eventually gives $T$.
Additionally,
$\pi$ is injective,
there is no $X \in \Gamma$ with $|\pi(X)| = 0$, 
and for each $X \in \Gamma \setminus \{X_T\}$, there is a $Y \in \Gamma$ such that 
$X$ is contained in $\pi(Y)$.
Obviously, $\BunpouT$ has no cycle.

For simplicity, we stipulate that $\pi(c) = c$ for $c \in \Sigma$.
We say that a non-terminal ($\in \Gamma$) or a character ($\in \Sigma$) is a \teigi{symbol}, 
and denote the set of characters and non-terminals with $\Symbols := \Sigma \cup \Gamma$.
We understand $\pi$ also as a string morphism $\pi : \Symbols^* \rightarrow \Symbols^*$ by applying $\pi$ on each symbol of the input string.
This allows us to define the \teigi{expansion}~$\pi^*(X)$ of a symbol~$X$, 
which is the iterative application of $\pi$ until obtaining a string of characters, i.e., $\pi^*(X) \subset \Sigma^*$ and $\pi^*(X_T) = T$.
Since $\pi(X)$ is deterministically defined, we use to say \emph{the right hand side of~$X$} for $\pi(X)$.
The lexicographic order on $\Sigma$ induces an ordering on $\Gamma$ by saying that $X \prec Y$ if and only if $\pi^*(X) \prec \pi^*(Y)$.

Further, we call an admissible grammar \teigi{factorizing} if we can split $\Gamma$ into the sets 
$\Takai{\Sigma}{1}, \ldots, \Takai{\Sigma}{t_T}$ such that
$\Takai{\Sigma}{t_T} = \{X_T\}$, 
and $\pi : \Takai{\Sigma}{h} \rightarrow \Takai{\Sigma}{h-1}$ with $\Takai{\Sigma}{0} = \Sigma$
is well-defined for each $h \in [1..t_T]$.
In particular, $\Takai{\pi}{t_T}(X_T) = T$. 
We say that $\BunpouT$ has the height~$t_T$, and that $\Takai{\Sigma}{h}$ are the non-terminals on height~$h$.
We write $\Takai{T}{h} = \Takai{\pi}{t_T-h}(X_T)$ with $\Takai{T}{0} = T$,
and $\Takai{\sigma}{h} := |\Takai{\Sigma}{h}|$ for $h \ge 0$.
Examples for factorizing grammars are ESP~\cite{cormode07esp} and HSP~\cite{fischer20deterministic}, but not Re-Pair~\cite{larsson99repair} or sequitur~\cite{nevill-manning97sequitur} in general.
Another example is GCIS, which we review next.

\subsection{Grammar Compression Based on Induced Suffix Sorting}
SAIS~\cite{nong11sais} is a linear-time algorithm for computing the suffix array~\cite{manber93sa}.
We briefly sketch the parts of SAIS needed for constructing the GCIS grammar.
Starting with a text~$T[1..n]$, we pad it with artificial characters \texttt{\#} and \texttt{\$} to its left and right ends, respectively,
such that $T[0] = \texttt{\#}$ and $T[n+1] = \texttt{\$}$.
We stipulate that $\texttt{\#} < \texttt{\$} < c$ for each character~$c \in \Sigma$.
Central to SAIS is the type assignment to each suffix, which is either~$\typeL$ or~$\typeS$:
\begin{itemize}
  \item  $T[i..]$ is an \typeL{} suffix if $T[i..] \succ T[i+1..]$, or
  \item  $T[i..]$ is an \typeS{} suffix otherwise, i.e., $T[i..] \prec T[i+1..]$,
\end{itemize}
where we stipulate that $T[n+1] = \texttt{\$}$ is always type~$\typeS$.
Since it is not possible that $T[i..] = T[i+1..]$, SAIS assigns each suffix a type.
An \typeS{} suffix $T[i..]$ is additionally an \typeHoshi{} suffix 
if $T[i-1..]$ is an \typeL{} suffix.
Note that $T[0..]$ is an \typeS{} suffix since $\texttt{\#}$ is the smallest character; we further let it be \typeHoshi{}.
The substring between two succeeding $\typeHoshi$ suffixes is called an \teigi{LMS substring}. 
In other words, a substring $T[i..j]$ with $i < j$ is an LMS substring if and only if $T[i..]$ and $T[j..]$ are \typeHoshi{} suffixes and 
there is no $k \in [i+1..j-1]$ such that $T[k..]$ is an \typeHoshi{} suffix.

The LMS substrings induce a factorization of $T[0..n+1] = T_1 \cdots T_t$, 
where each factor starts with an LMS substring.
We call this factorization \teigi{LMS factorization}.
By replacing each factor~$T_x$ by the lexicographic rank of its respective LMS substring\footnote{Note that SAIS uses a ordering different to the lexicographic order.
However, the lexicographic order is sufficient for the computation of the grammar.},
we obtain a string $\Takai{T}{1}$ of these ranks.
We recurse on $\Takai{T}{1}$ until we obtain a string $\Takai{T}{t_T-1}$ whose rank-characters are all unique or whose LMS factorization consists of at most two factors.
If we, instead of assigning ranks, assign each LMS substring a non-terminal, and recurse on a string of non-terminals, 
we obtain a grammar \BunpouT{} that is factorizing. 
Specifically, the right hand side of a non-terminal is an LMS substring without its last character, and the special characters $\texttt{\#}$ and $\texttt{\$}$ are omitted.
The start symbol is defined by $X_T \rightarrow \Takai{T}{t_T}$.

\begin{lemma}[\cite{nunes18grammar}]\label{lemGCISconstruct}
  The GCIS grammar \BunpouT{} can be constructed in \Oh{n} time.
  $\BunpouT$ is \emph{reduced}, meaning that we can reach all non-terminals of $\Gamma$ from $\Takai{X}{t_T}$.
\end{lemma}

Since there are no neighboring \typeHoshi{} suffixes, an LMS substring has a length of at least three, and therefore the right-hand sides of all non-terminals are of length at least two (except maybe for the first factor).
This means that the length of $\Takai{T}{i}$ is at most half of the length of $\Takai{T}{i-1}$ for $i \ge 1$.
Consequently, the height $t_T$ is \Oh{\lg n}.

\subsection{Example for a GCIS grammar}\label{secExampleGCIS}
We build GCIS on the example text $T := \texttt{bacabacaacbcbc}$.
For that, we determine the types of all suffixes, which determine the LMS substrings, as shown in  \cref{figGCIStext}.

\begin{figure}[H]
  \centering{\begin{tikzpicture}[global scale=0.85]
\matrix[array,
	, row 1/.style={}
	, row 2/.style={nodes={font=\ttfamily,fill=none, minimum size=0mm, text height=0.33em}}
	, row 3/.style={nodes={font=\ttfamily\small,fill=none, minimum size=0mm, text height=0.33em}}
] (text) {0  & 1 & 2  & 3 & 4  & 5 & 6  & 7 & 8  & 9 & 10 & 11 & 12 & 13 & 14 & 15 \\
    \# & b & a  & c & a  & b & a  & c & a  & a & c  & b  & c  & b  & c  & \$ \\
    S* & L & S* & L & S* & L & S* & L & S* & S & L  & S* & L  & S* & L  & S* \\
	};
\node [left=0em of text-2-1,text height=0.33em] {\texttt{\$}$T$\texttt{\#} = };

	\begin{scope}[transform canvas={yshift=-2em}]
	  \node[fit=(text-3-1)(text-3-3)]{\texttt{D}};
	  \node[fit=(text-3-3)(text-3-5)]{\texttt{C}};
	  \node[fit=(text-3-5)(text-3-7)]{\texttt{B}};
	  \node[fit=(text-3-7)(text-3-9)]{\texttt{C}};
	  \node[fit=(text-3-9)(text-3-12) ]{\texttt{A}};
	  \node[fit=(text-3-12)(text-3-14)]{\texttt{E}};
	  \node[fit=(text-3-14)(text-3-16)]{\texttt{E}};
    \end{scope}

\LMSbracket{1}{3}
\LMSbracket{3}{5}
\LMSbracket{5}{7}
\LMSbracket{7}{9}
\LMSbracket{9}{12}
\LMSbracket{12}{14}
\LMSbracket{14}{16}
\end{tikzpicture}
\vspace{1em}
  }\caption{Application of GCIS on the text $T := \texttt{bacabacaacbcbc}$.
    The type of each suffix is shown below its starting position.
The rectangular bracket below the types demarcate the LMS substrings.
  }
  \label{figGCIStext}
\end{figure}

We obtain the grammar~$\BunpouT$ with the following rules:
$\texttt{A} \rightarrow \texttt{aac}$,
$\texttt{B} \rightarrow \texttt{ab}$,
$\texttt{C} \rightarrow \texttt{ac}$,
$\texttt{D} \rightarrow \texttt{b}$, and
$\texttt{E} \rightarrow \texttt{bc}$.
The grammar has $\Takai{\sigma}{1} := 5$ non-terminals on height~$1$.
By replacing the LMS substrings with the respective non-terminals, we obtain the string
$\Takai{T}{1} := \texttt{DCBCAEE}$.
Since there are two occurrences of \texttt{E}, we would recurse, but here, and in the following examples, we stop at height~$1$ for simplicity.
In what follows, we study an approach that builds the BWT on this text, which is given by $\Takai{\BWT}{1} := \texttt{ECCBD\$EA}$.

\section{FM-Indexing the GCIS Grammar}\label{secFactContext}
The main idea of our approach is that we build the GCIS grammar \BunpouP{} on $P$ and
translate the matching problem of $P$ in $T$ to matching $\Takai{P}{h}$ in $\Takai{T}{h}$, for a height $h \in [1..\min(t_T,t_P)-1]$,
with $t_P$ being the height of \BunpouP{}.
The problem is that the LMS factorization of $P$ and the LMS factorization of the occurrences of $P$ in $T$ can look differently
since the occurrences of $P$ in $T$ are not surrounded by the artificial characters \texttt{\#} and \texttt{\$}, but by different contexts of $T$.
The question is whether there is a substring of $\Takai{P}{h}$, for which we can be sure that each occurrence of $P$ in $T$ is represented in $\Takai{T}{h}$ by a substring containing $\Takai{P}{h}$. 
We call such a maximal substring a \teigi{core}, and give a characterization similar to \citet[Section 4.1]{akagi21grammar} that determines this core:

\subsection{Cores}\label{secCores}
Given a pattern $P[1..m]$, we pad it like the text with the artificial characters \texttt{\#} and \texttt{\$}, and compute its LMS factorization.
Now, we study the change of the LMS factorization when prepending or appending characters to $P$, i.e.,
we change $P$ to $\texttt{c} P$ or $P \texttt{c}$ for a character~$\texttt{c} \in \Sigma$, 
while keeping the artificial characters \texttt{\#} and \texttt{\$} at the left and right ends, respectively.
We claim that
(a) prepending characters can only extend the leftmost factor or let a new factor emerge consisting only of the newly introduced character, and
(b) appending characters can split the last factor at the beginning of the rightmost character run into two.
Consequently, given that the LMS factorization of $P$ is $P = P_1 \cdots P_p$, fix an occurrence of $P$ in $T$.
Then this occurrence is contained in the LMS factors $P'_1 P_2 \cdots P_{p-1} P'_p P'_{p+1}$, where $P_1$ is a (not necessarily proper) suffix of $P'_1$, 
and either
(a) $P'_{p+1}$ is empty and $P_p$ is a (not necessarily proper) prefix of $P'_p$, or
(b) $P'_p$ is $P_p$ without its last character run, which is the prefix of $P'_{p+1}$.

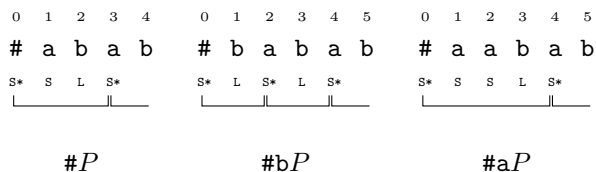
\begin{figure}
    \centering{\tikzstyle{prepend} = [array,
	, row 2/.style={nodes={font=\ttfamily, fill=none, minimum size=0mm, text height=0.5em}}
	, row 3/.style={nodes={font=\tiny\ttfamily,fill=none, minimum size=0mm, text height=0.33em}}
	, row 1/.style={nodes={font=\tiny,fill=none, minimum size=0mm, text height=0.33em}}
	]
	\newcommand*{\LMSleftbracket}[2]{\begin{scope}[transform canvas={yshift=-0.4em}]
		\draw  ([yshift=0.4em,xshift=-0.1em]text-3-#1.south) -- ([xshift=-0.1em]text-3-#1.south) -- ([xshift=0.1em]text-3-#2.south) {};
	\end{scope}
    }
\begin{tikzpicture}
\matrix[prepend] (text) {0 & 1 & 2 & 3 & 4 \\
    \# & a & b  & a & b \\
    S* & S & L  & S* &  \\
	};
	\node [below=1.5em of text] {\texttt{\#}$P$};
\LMSbracket{1}{4}
	\LMSleftbracket{4}{5}
\end{tikzpicture}
\begin{tikzpicture}
\matrix[prepend] (text) {0 & 1 & 2 & 3 & 4 & 5 \\
    \# & b & a & b  & a & b \\
    S* & L & S* & L  & S* &  \\
	};
	\node [below=1.5em of text] {$\texttt{\#b}P$};
\LMSbracket{1}{3}
\LMSbracket{3}{5}
	\LMSleftbracket{5}{6}
\end{tikzpicture}
\begin{tikzpicture}
\matrix[prepend] (text) {0 & 1 & 2 & 3 & 4 & 5 \\
    \# & a & a & b  & a & b \\
    S* & S & S & L  & S* &  \\
	};
	\node [below=1.5em of text] {$\texttt{\#a}P$};
\LMSbracket{1}{5}
	\LMSleftbracket{5}{6}
\end{tikzpicture}
    }\caption{Prepending one of the characters \texttt{a} or \texttt{b} to $P = \texttt{abab}$. 
    The rectangular brackets demarcate the LMS substrings. The two cases are studied in \cref{secCores}.}
    \label{figPrepending}
\end{figure}

\paragraph{Prepending}
Suppose we prepend a new character~$\texttt{c}$ to $P$ such that we get $P' := \texttt{c} P$ with $P'[1] = \texttt{c}$ and $P'[0] = \texttt{\#}$.
Then none of the types changes, i.e., the type of $P'[i+1]$ is the type of $P[i]$ for $i \ge 1$, since the type of a suffix is independent of its preceding suffixes.
It is left to determine the type of $P'[1]$ and to update the first LMS substring of~$P'$ (cf.~\cref{figPrepending}):
If $P[1]$ is type~$\typeS$ but $P'[2] (= P[1])$ has become type~$\typeHoshi$ ($P'[1] > P'[2]$), 
then we introduce a new LMS substring $P'[0..2]$ and let the old LMS substring formerly covering $P[0]$ and $P[1]$ start at $P'[2]$.
Otherwise, we extend the leftmost factor.

\paragraph{Appending}
Let us fix an occurrence of the pattern~$P$ in the text~$T$, let $m'$ be the position in $T$ matching $P[m]$, and assume that the LMS factorization of~$P$ is $P = P_1 \ldots P_p$ with $p > 2$.
Note that $P[m]$ is always $\typeL$ since its successor is \texttt{\$}.
Given the last two factors of $P$ are $P_{p-1}$ and $P_p$, we have two cases to consider of how the LMS factor in $T$ covering the same characters as $P_{p-1}$ and $P_p$ look like.
First, suppose that $P[m-1]$ is \typeS{}. 
Then $P[m]$ is contained in $P_p$. 
Regardless of the type of $T[m'-1]$, the text factor~$F$ covering $T[m']$ has $P_p$ as a (not necessarily proper) prefix,
and its preceding factor is $P_{p-1}$ (assuming that $p > 2$).
Second, suppose that $P[m-1]$ is \typeL{}.
If $T[m']$ is \typeL{}, then we have the same setting as above (we do not introduce a new LMS substring with an extra \typeHoshi{} suffix).
However, if $T[m']$ is \typeS{}, then the factorization of $P$'s occurrence in $T$ differs:
Let $\ell \ge 0$ be the largest value for which $P[m-\ell] = P[m-\ell-1] \ldots = P[m]$. 
Then $P[m-\ell],\ldots,P[m]$ are \typeL{} while $T[m'-\ell],\ldots,T[m']$ are \typeS{} with $T[m'-\ell]$ being \typeHoshi{}.
Since $P[m-\ell],\ldots,P[m]$ are contained in $P_p$, the text factor covering $T[m'-\ell-1]$ is a prefix of $P_p$, and its preceding factor is equal to $P_{p-1}$.
In total, when matching the last LMS factors of $P$ with the occurrences of $P$ in $T$, only the last character run in $P$ can be contained in a different LMS factors.
\Cref{figDanglingCharacter} visualizes our observation considering the additional case that $P[m-1]$ is \typeHoshi{}, which is covered in our first case.

\begin{figure*}
  \centering

\newcommand{\ZeichneText}[1]{

	\node [font=\scriptsize] at ([yshift=0.4em]mat-1-#1.north) {$T[m']$};
	\draw (mat-1-1.north west) rectangle (mat.south east |- mat-1-1.south);
	\node [left of=mat-1-1] {$T = $};
}

	\newcounter{index}
	\newcounter{lastposition}
	\newcommand{\ZeichnePattern}[1]{\pgfmathparse{\mypatterncols-1}
	\setcounter{lastposition}{\pgfmathresult}

	\node [font=\scriptsize] at ([yshift=0.4em]pattern-1-\thelastposition.north) {$P[m]$};

\node [left of=pattern-1-1] {$P = $};
	\draw (pattern-1-1.north west) rectangle ([xshift=0.1em]pattern-1-\thelastposition.south east |- pattern-1-1.south);
\ifthenelse{#1 = 0}{\setcounter{index}{1}
}{\node [fit=(pattern-2-1.south) (pattern-1-#1.north), Block, color=solarizedRed, label={[color=solarizedRed]below:$P_{p-1}$}] {};
	\pgfmathparse{#1+1}
	\setcounter{index}{\pgfmathresult}
}
	\node [fit=(pattern-2-\theindex.south) (pattern-1-\thelastposition.north), Block, draw=solarizedBlue, label={[color=solarizedBlue]below:$P_{p}$}] {};
}

\begin{tikzpicture}
\matrix[Pattern] (pattern) {$\cdots$ & a & b  & c & \# \\
	     & S & S  & L & S* \\
	};
\ZeichnePattern{0}
\node [below=1.5em of pattern] {1. $P[m-1]$ is \typeS{}.};
\end{tikzpicture}
\begin{tikzpicture}
\matrix[Text] (mat) {$\cdots$ & a & b  & c & d & c & b & a & $\cdots$ \\
          & S & S  & S & L & L & L & \\
	};
	\node [fit=(mat-2-1.south) (mat-1-\mytextcols.north), Block, draw=solarizedBlue, label={[color=solarizedBlue]below:$P'_p$}] {};
	\ZeichneText{4}
\node [below=1.5em of mat] {1.a $T[m']$ is \typeS{}.};
\end{tikzpicture}
\begin{tikzpicture}
\matrix[Text] (mat) {$\cdots$  & a & b & c & b & a  & b & c & $\cdots$ \\
		   & S & S & L & L & S* & S & L & \\
	};
	\ZeichneText{4}
	\node [fit=(mat-2-1.south) (mat-1-5.north), Block, draw=solarizedBlue, label={[color=solarizedBlue]below:$P'_p$}] {};
	\node [transform canvas={yshift=-0.05em},fit=(mat-2-6.south) (mat-1-\mytextcols.north), Block, draw=solarizedGreen, label={[color=solarizedGreen,yshift=-0.05em]below:$P'_{p+1}$}] {};
\node [below=1.5em of mat] {1.b $T[m']$ is \typeL{}.};
\end{tikzpicture}

\begin{tikzpicture}
\matrix[Pattern] (pattern) {$\cdots$ & c & b  & c & \# \\
	     & L & S* & L  & S* \\
	};
\ZeichnePattern{2}
\node [below=1.5em of pattern] {2. $P[m-1]$ is \typeHoshi{}.};
\end{tikzpicture}
\begin{tikzpicture}
\matrix[Text] (mat) {$\cdots$ & c & b  & c & d & c & b & a & $\cdots$ \\
          & L & S* & S & L & L & L & \\
	};
	\ZeichneText{4}
	\node [fit=(mat-2-1.south) (mat-1-2.north), Block, draw=solarizedRed, label={[color=solarizedRed]below:$P_{p-1}$}] {};
	\node [transform canvas={yshift=-0.05em},fit=(mat-2-3.south) (mat-1-\mymatcols.north), Block, draw=solarizedBlue, label={[color=solarizedBlue,yshift=-0.05em]below:$P'_{p}$}] {};
\node [below=1.5em of mat] {2.a $T[m']$ is \typeS{}.};
\end{tikzpicture}
\begin{tikzpicture}
\matrix[array] (mat) {$\cdots$ & c & b  & c & b & a & b & c & $\cdots$ \\
		  & L & S* & L & L & S* & S & L  & \\
	};
	\ZeichneText{4}
	\node [fit=(mat-2-1.south) (mat-1-2.north), Block, draw=solarizedRed, label={[color=solarizedRed]below:$P_{p-1}$}] {};
	\node [transform canvas={yshift=-0.08em},fit=(mat-2-3.south) (mat-1-5.north), Block, draw=solarizedBlue, label={[color=solarizedBlue,yshift=-0.08em]below:$P'_{p}$}] {};
	\node [transform canvas={yshift=-0.05em},fit=(mat-2-6.south) (mat-1-9.north), Block, draw=solarizedGreen, label={[color=solarizedGreen,yshift=-0.05em]below:$P'_{p+1}$}] {};
\node [below=1.5em of mat] {2.b $T[m']$ is \typeL{}.};
\end{tikzpicture}

\begin{tikzpicture}
\matrix[Pattern] (pattern) {$\cdots$ & d  & c & \# \\
	     & L  & L & S* \\
	};
\ZeichnePattern{0}
\node [below=1.5em of pattern] {3. $P[m-1]$ is \typeL{}.};
\end{tikzpicture}
\begin{tikzpicture}
\matrix[Text] (mat) {$\cdots$ & d & c  & d & c & b & a & $\cdots$ \\
          & L & S* & L & L & L & & \\
	};
	\ZeichneText{3}
	\node [fit=(mat-2-1.south) (mat-1-2.north), Block, draw=solarizedBlue, label={[color=solarizedBlue]below:$P'_p$}] {};
	\node [transform canvas={yshift=-0.05em},fit=(mat-2-3.south) (mat-1-\mymatcols.north), Block, draw=solarizedGreen, label={[color=solarizedGreen,yshift=-0.05em]below:$P'_{p+1}$}] {};
\node [below=1.5em of mat] {3.a $T[m']$ is \typeHoshi{}.};
\end{tikzpicture}
\begin{tikzpicture}
\matrix[array] (mat) {$\cdots$  & d & c  & b & a  & b & $\cdots$ \\
		   & L & L & L & S* &   & \\
	};
	\ZeichneText{3}
	\node [fit=(mat-2-1.south) (mat-1-4.north), Block, draw=solarizedBlue, label={[color=solarizedBlue]below:$P'_p$}] {};
	\node [transform canvas={yshift=-0.05em},fit=(mat-2-5.south) (mat-1-\mymatcols.north), Block, draw=solarizedGreen, label={[color=solarizedGreen,yshift=-0.05em]below:$P'_{p+1}$}] {};
\node [below=1.5em of mat] {3.b $T[m']$ is \typeL{}.};
\end{tikzpicture}
\caption{Difference in the factorization of the pattern and its occurrences in the text.
  Let $T[m']$ be the $m$-th position of an occurrence of $P$ in $T$.
    Only the last character run of the LMS factorization of $P$ can be found in a different factor when considering an occurrence of $P$ in $T$ as part of the LMS factorization of~$T$.
In Cases~1 and~2, we extend the last factor~$P_p$, while we split~$P_p$ in Case~3, moving its last character to a new factor~$P'_{p+1}$.}
	\label{figDanglingCharacter}
\end{figure*}

\subsection{Pattern Matching}\label{secPatternMatching}
For simplicity, assume that we stop the grammar construction on the first level, i.e., 
after computing the factorization of the plain text
such that $t_T = 2$.
We additionally build the BWT on $\Takai{T}{t_T-1}$ and call it \Takai{\BWT}{t_T-1}. 
It can be computed in linear time by using an (alphabet-independent) linear-time suffix array construction algorithm like SAIS\@.

Now, given a pattern~$P$, we compute the GCIS grammar~$\BunpouP$ on $P$, 
where we use the same non-terminals as in $\BunpouT$ whenever their right hand sides match.
Then there are non-terminals $Y_1, \ldots, Y_p$ such that $P$ has the LMS factorization $P = P_1 \cdots P_p$ with $P_y = \pi(Y_y)$ for each $y \in [1..p]$.
According to \cref{secCores} each occurrence of $P$ in $T$ is captured by an occurrence of $Y_{2} \cdots Y_{p-1}$ in $\Takai{T}{1}$.
So $Y_2, \ldots, Y_{p-1}$ do not only appear as non-terminals in the grammar of~$T$, but they also appear as substrings in $\Takai{T}{1}$ (if $P$ occurs in~$T$).
In what follows, we call $Y_2, \ldots, Y_{p-1}$ the \teigi{core} of $P$, and show how to use the core to find $P$ 
via \Takai{\BWT}{1} and a dictionary on right hand sides of the non-terminals of $\BunpouT$.

If we turn \Takai{\BWT}{1} into an FM-index by representing it by a wavelet tree, 
it can find the core of $P$ in $p-2$ backward search steps,
i.e., returning an interval in the BWT that corresponds to all occurrences of $Y_2 \cdots Y_{p-1}$ in $\Takai{T}{1}$,
which corresponds to all occurrences of $P_2 \cdots P_{p-1}$ in $T$.
We can extend this interval to an interval covering all occurrences $P_1 \cdots P_{p-1}$ with the following trick:
On constructing the wavelet tree on \Takai{\BWT}{1},
we encode the symbols of $\Takai{T}{1}$ by the colexicographic order of their right hand sides. 
See \cref{tabColexRank} for the colexicographic ranking of the non-terminals, and \cref{figWaveletTree} for the wavelet tree of our running example.
To understand our modification, we briefly review the wavelet tree under that aspect:
The wavelet tree is a binary tree.
The root node stores for each text position~$i$ of \Takai{\BWT}{1} a bit for whether the colexicographic rank of this $\Takai{\BWT}{1}[i]$ is larger than $\Takai{\sigma}{1}/2$.
Its left and right children inherit the input string omitting the marked and unmarked positions, respectively such that the left and the right children obtain strings whose symbols have colexicographic ranks in $[0..\Takai{\sigma}{1}/2]$ and
$[\Takai{\sigma}{1}/2+1..\Takai{\sigma}{1}]$, respectively. The construction works then recursively in that the children themselves create bit vectors to partition the symbols. The recursion ends whenever a node receives a unary string.

By having ranked the non-terminals ($\in \Takai{\Sigma}{1}$) colexicographically during the construction of the wavelet tree of the BWT,
matching $\pi(Y_1)$ is done by a top-down traversal of the wavelet tree, starting at the root.
By doing so, we can find the lowest node whose leaves represent the positions of all non-terminals having $\pi(Y_1)$ as a suffix, 
within the query range of $\pi(Y_2) \cdots \pi(Y_{p-1})$.

\begin{table}
  \centering
  \caption{Colexiographic ranking of the non-terminals of \cref{secExampleGCIS}. We additionally add the artificial character \texttt{\$} with rank~$0$ because it is later used in \Takai{\BWT}{1}.}
  \label{tabColexRank}
		\begin{tabular}{crc}
	  \toprule
	  $X$ & $\pi(X)$ & colex.\ rank \\
	  \midrule
	  \texttt{A} & \texttt{aac} & 4 \\
	  \texttt{B} & \texttt{~ab} & 2 \\
	  \texttt{C} & \texttt{~ac} & 3 \\
	  \texttt{D} & \texttt{~~b} & 1 \\
	  \texttt{E} & \texttt{~bc} & 5 \\
	  \texttt{\$} & \texttt{~~\$} & 0 \\
\bottomrule
	\end{tabular}
\end{table}

Finally, it is left to find the missing suffix.
Let $\mathcal{R} := \{R_k\}_{k} \subset \Takai{\Sigma}{1}$ be the set of all rules~$R_k$ with $P_p$ being a (not necessarily proper) prefix of $\pi(R_k)$.
Since each $R_k \in \mathcal{R}$ received a rank according to the lexicographic order of its right hand side, 
the elements in $\mathcal{R}$ form a consecutive interval in \BWT{}, and this interval corresponds to occurrences of $P_p$.
So staring with this interval the aforementioned backward search gives us occurrences of $P$.

However, the final range may not contain \emph{all} occurrences.
That is because, according to \cref{secCores}, the rightmost non-terminal may not cover $P_p$ completely, but only
$P_p[1..|P_p|-\ell-1]$, 
where $P_p[|P_p|-\ell..|P_p|]$ is the longest character run that is a suffix of $P_p$, for $\ell \ge 0$.
Now, suppose that the rule $X_p \rightarrow P_p[1..|P_p|-\ell]$ exists,
then we need to check, for all non-terminals in the set $\mathcal{U} = \{U_j\}_j$ with $P_p[|P_p|-\ell..|P_p|]$ being a prefix of $\pi(U_j)$,
whether $X_p U_j$ is a substring of $T$. 
With analogous reasoning, the occurrences of all elements of $\mathcal{U} \subset \Takai{\Sigma}{1}$ form a consecutive range in \BWT{},
and with a backward search for $X_p$ we obtain another range corresponding to $P_p$.
However, this range combined with the range for $\mathcal{R}$ gives \emph{all} occurrences of $P_p$.
Consequently, if $X_p$ exists, we need to perform the backward search not only for the range of $\mathcal{R}$, but also for $X_p \mathcal{U}$.

\subsection{Example for Pattern Matching}
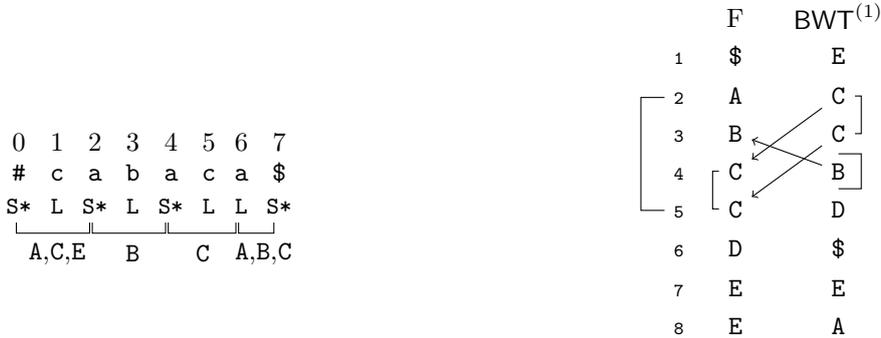
\begin{figure}[t]
  \centering{\adjustbox{valign=c}{\begin{tikzpicture}
\matrix[array,
	, row 1/.style={}
	, row 2/.style={nodes={font=\ttfamily,fill=none, minimum size=0mm, text height=0.33em}}
	, row 3/.style={nodes={font=\ttfamily\small,fill=none, minimum size=0mm, text height=0.33em}}
	] (text) {0  & 1 & 2  & 3 & 4  & 5 & 6  & 7  \\
    \# & c & a  & b & a  & c & a  & \$ \\
    S* & L & S* & L & S* & L & L & S*  \\
	};

	\begin{scope}[transform canvas={yshift=-2em}]
	  \node[fit=(text-3-1)(text-3-3)]{\texttt{A},\texttt{C},\texttt{E}};
	  \node[fit=(text-3-3)(text-3-5)]{\texttt{B}};
	  \node[fit=(text-3-5)(text-3-7)]{\texttt{C}};
	  \node[fit=(text-3-7)(text-3-8)]{\texttt{A},\texttt{B},\texttt{C}};
    \end{scope}

\LMSbracket{1}{3}
\LMSbracket{3}{5}
\LMSbracket{5}{7}
\LMSbracket{7}{8}
\end{tikzpicture}
}\hfill\adjustbox{valign=c}{\begin{tikzpicture}
\usetikzlibrary{calc}

    \tikzset{
    ncbar angle/.initial=90,
    ncbar/.style={
        to path=(\tikztostart)
        -- ($(\tikztostart)!#1!\pgfkeysvalueof{/tikz/ncbar angle}:(\tikztotarget)$)
        -- ($(\tikztotarget)!($(\tikztostart)!#1!\pgfkeysvalueof{/tikz/ncbar angle}:(\tikztotarget)$)!\pgfkeysvalueof{/tikz/ncbar angle}:(\tikztostart)$)
        -- (\tikztotarget)
    },
    ncbar/.default=0.5cm,
}
\matrix[matrix of nodes
    , nodes in empty cells
	, column sep=1em
	, row sep=0.5mm
	, font=\ttfamily
	, column 1/.style={nodes={font=\ttfamily\scriptsize, fill=none, minimum size=0mm, text height=0.5em, column sep=0em}}
] (fm) {1 &  \$ & & E \\
    2 &  A & & C \\
    3 &  B & & C \\
    4 &  C & & B \\
    5 &  C & & D \\
    6 &  D & & \$ \\
    7 &  E & & E \\
    8 &  E& & A \\
	};

	\node [above=0cm of fm-1-4] {$\Takai{\BWT}{1}$};
	\node [above=0cm of fm-1-2] {F};
	\draw (fm-2-1) to [ncbar=-0.5cm] (fm-5-1);
	\draw (fm-2-4) to [ncbar=0.3cm]  (fm-3-4);
	\draw (fm-4-2) to [ncbar=-0.3cm]  (fm-5-2);
	\draw (fm-4-4.north) to [ncbar=0.3cm]  (fm-4-4.south);
\draw [->] (fm-2-4) -- (fm-4-2);
\draw [->] (fm-3-4) -- (fm-5-2);
\draw [->] (fm-4-4) -- (fm-3-2);

\end{tikzpicture}
}
  }\caption{Matching the pattern $P = \texttt{cabaca}$ in $\Takai{\BWT}{1}$ built on~$\Takai{T}{1} := \texttt{DCBCAEE}$ with the algorithm described in \cref{secPatternMatching}.
    \emph{Left:} Application of GCIS on $P$, analogously to \cref{figGCIStext} for the text. While we can determine the non-terminals corresponding to $P_2, P_3,$ and $P_4$, we have several candidate non-terminals that have $P_1$ and $P_4$ as a suffix or prefix, respectively, which we list below the brackets demarcating the LMS substrings of~$P$.
    \emph{Right} Matching $P_2 P_3 P_4$ in $\Takai{\BWT}{1}$ with the backward search.
}
  \label{figPatternMatch}
\end{figure}

Continuing with \cref{secExampleGCIS}, let 
$P := \texttt{cabaca}$ be a given pattern.
We obtain the factorization of $P$ with its core \texttt{BC} as shown in \cref{figPatternMatch} on the left.
The pattern is divided into four factors $P_1$, $P_2$, $P_3$, and $P_4$, where we know that $P_2$ and $P_3$ are the right hand sides of \texttt{B} and \texttt{C}, respectively.
We find that only the non-terminals \texttt{A}, \texttt{B}, and \texttt{C} have $P_4 = \texttt{a}$ as a prefix of their right hand sides.
These form a consecutive interval $[2..5]$ in $\Takai{\BWT}{1}$. 
With the backward search, we can find the interval of $P_2 P_3 P_4$ from $[2..5]$, as shown in the right of \cref{figPatternMatch}:
From $[2..5]$, we match $P_3$ corresponding to \texttt{C}, which gives the first and the second \texttt{C} in $F$, 
represented by the interval $[4..5]$. From there, we match $P_2$ corresponding to \texttt{B}, which gives the first \texttt{B} at position~$3$. 

\begin{figure}[t]
  \centering{\begin{tikzpicture}
\tikzstyle{wavelet} = [array,
	, row 2/.style={nodes={font=\ttfamily, fill=none, minimum size=0mm, text height=0.5em}}
	, row 3/.style={nodes={font=\ttfamily,fill=none, minimum size=0mm, text height=0.33em}}
	, row 1/.style={nodes={font=\tiny,fill=none, minimum size=0mm, text height=0.33em}}
	]

\matrix[wavelet
] (wt0) {5 & 3 & 3 & 2 & 1 & 0  & 5 & 4 \\
  E & C & C & B & D & \$ & E & A \\
  1 & 1 & 1 & 0 & 0 & 0  & 1 & 1 \\
	};
	\node [left=0cm of wt0-2-1] {$\Takai{\BWT}{1}$ = };

\matrix[wavelet, below=1cm of wt0.west
] (wtb) {2 & 1 & 0  \\
  B & D & \$ \\
  1 & 0 & 0  \\
	};

\matrix[wavelet, below=1cm of wtb.west
] (wtba) {1 & 0  \\
  D & \$ \\
  1 & 0  \\
	};

      \node [below=1cm of wtb.east] (B) {\texttt{B}};

\matrix[wavelet, below=1cm of wt0.east
] (wtc) {5 & 3 & 3 & 5 & 4 \\
  E & C & C & E & A \\
  1 & 0 & 0 & 1 & 0 \\
	};

\matrix[wavelet, below=1cm of wtc.west
] (wtca) {3 & 3 & 4 \\
  C & C & A \\
  0 & 0 & 1 \\
	};

      \node [below=1cm of wtc.east] (E) {\texttt{E}};

\draw (wt0) -- node [midway,anchor=east] {\texttt{b\$}} (wtb);
\draw (wt0) -- node [midway,anchor=west] {\texttt{c}} (wtc);
\draw (wtc) -- node [midway,anchor=east] {\texttt{a}} (wtca);
\draw (wtc) -- node [midway,anchor=west] {\texttt{b}} (E);
\draw (wtb) -- node [midway,anchor=east] {} (wtba);
\draw (wtb) -- node [midway,anchor=west] {\texttt{a}} (B);

\end{tikzpicture}
  }\caption{The wavelet tree of $\Takai{\BWT}{1}$ on our running example.
The wavelet tree ranks each non-terminal by the colexicographic rank of its right hand side.
Each row of the wavelet tree is depicted as a small matrix, where the actual data is the last row.
The first row of each matrix consists of the ranks and the second row consists of the corresponding characters ($\in \Takai{\Sigma}{1}$).
An edge on the $i$-th level lists all possible starting characters of the $i$-th suffix of the right hand sides of all non-terminals below this edge.
  }
  \label{figWaveletTree}
\end{figure}

To match further, we look at the wavelet tree given in \cref{figWaveletTree}.
There, we can use the edges to match the non-terminals with a pattern \emph{backwards}.
For instance, all non-terminals having $P_1 = \texttt{c}$ as a suffix are found in the right subtree of the root.
However, we are interested in completing the range of~$P$ from the range of~$P_2 P_3 P_4$, which consists of the single position~$2$.
Hence, we look for all non-terminals having $P_1$ as a suffix \emph{within this range}, 
which gives us the second \texttt{C}.

Finally, we explain our dictionary used for finding the non-terminals based on their right hand sides.
This dictionary is represented by a trie, and implemented by the extended Burrows-Wheeler Transform (XBWT)~\cite{ferragina09xbwt}.
We use the XBWT because it
supports substring queries~\cite{manzini16xbwt}, which allow us to extend a substring match by appending or prepending characters to the query.

\subsection{XBWT}\label{secXBWT}
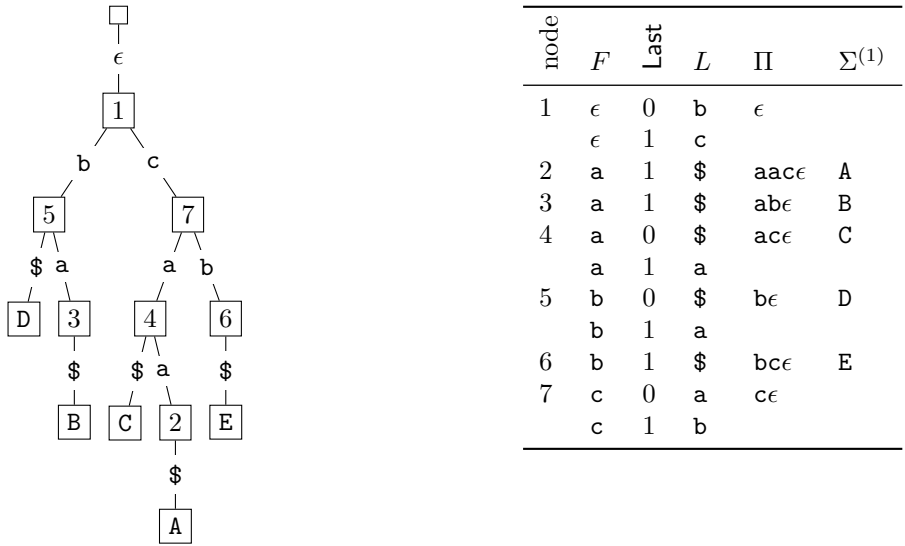
\begin{figure}[t]
  \centering{\adjustbox{valign=t}{\tikzstyle{lab} = [midway,fill=white,font=\ttfamily]
\begin{forest}
  for tree={draw,l sep+=1.4em}
[
[1,edge label={node[lab]{$\epsilon$}}
[5,edge label={node[lab]{b}}
[\texttt{D},edge label={node[lab]{\$}}]
[3,edge label={node[lab]{a}}
[\texttt{B},edge label={node[lab]{\$}}]]]
[7,edge label={node[lab]{c}}
[4,edge label={node[lab]{a}}
[\texttt{C},edge label={node[lab]{\$}}]
[2,edge label={node[lab]{a}}
[\texttt{A},edge label={node[lab]{\$}}]]]
[6,edge label={node[lab]{b}}
[\texttt{E},edge label={node[lab]{\$}}]]
]]]
\end{forest}
}\hfill\adjustbox{valign=t}{\begin{tabular}{llllll}
	  \toprule
	  \rotatebox{90}{node} &	  $F$ & \rotatebox{90}{\Last{}} & $L$  & $\Pi$                     & \Takai{\Sigma}{1} \\
	  \midrule
			     1 &  $\epsilon$           & 0 & \texttt{b } & $\epsilon$ \\
			       &  $\epsilon$           & 1 & \texttt{c } & \\
			     2 &  \texttt{a}           & 1 & \texttt{\$} & \texttt{aac}$\epsilon$ & \texttt{A} \\
			     3 &  \texttt{a}           & 1 & \texttt{\$} & \texttt{ab}$\epsilon$  & \texttt{B} \\
			     4 &  \texttt{a}           & 0 & \texttt{\$} & \texttt{ac}$\epsilon$  & \texttt{C} \\
			       &  \texttt{a}           & 1 & \texttt{a } & \\
			     5 &  \texttt{b}           & 0 & \texttt{\$} & \texttt{b}$\epsilon$  & \texttt{D} \\
			       &  \texttt{b}           & 1 & \texttt{a } & \\
			     6 &  \texttt{b}           & 1 & \texttt{\$} & \texttt{bc}$\epsilon$ & \texttt{E} \\
			     7 &  \texttt{c}           & 0 & \texttt{a } & \texttt{c}$\epsilon$  \\
			       &  \texttt{c}           & 1 & \texttt{b} &  \\
\bottomrule
	\end{tabular}
 }
  }\caption{The trie (\emph{left}) on the reversed right hand side rules of all non-terminals of \Takai{\Sigma}{1},
    and its XBWT representation (\emph{right}), cf.~\cref{secXBWT}.
    The leaves are represented by the \texttt{\$} entries in $L$.
    The column \Takai{\Sigma}{1} gives the non-terminal associated with a leaf.
    Reading the leaves representing the non-terminals from left to right gives their colexiographic ranking, cf.~\cref{tabColexRank}.
    Each node is represented by as many rows as it has children.
}
  \label{figXBWT}
\end{figure}

The \teigi{grammar trie} of $\BunpouT$ on height~$h$ stores the reversed of the right hand sides of each non-terminal in $\Takai{\Sigma}{h}$ for $h \ge 1$, 
appended with an additional delimiter~$\texttt{\$} \not\in \Sigma$ smaller than all symbols.
Each leaf of the trie corresponds to a non-terminal. 
The trie for our running example is depicted on the left side of \cref{figXBWT}.
There, we additionally added an imaginary node as the parent of the root connected with an artificial character $\epsilon < \texttt{\$}$, 
which is needed for the XBWT construction.
The XBWT~\cite{ferragina09xbwt} of this trie is shown on the right of \cref{figXBWT}.
It consists of the arrays~$F$, \Last{}, and $L$; the other columns in the figure like~$\Pi$ are only for didactic reasons:
$L$ and $\Pi$ represent the labels of the paths from each trie node up to the root, where $L$ stores the first symbol, $\Pi$ stores the remaining part, and $F$ stores the first symbols of each string stored in $\Pi$.
Consequently, concatenating $L[i]$ and $\Pi[i]$ gives the path from a node to the root in the trie. 
Each pair $(L[i],\Pi[i])$ is permuted such that $\Pi$ is sorted lexicographically. 
  The last element with the same string in $\Pi$ is marked with a \bsq{1} in the bit vector~\Last{}.
  $L$ is represented with a wavelet tree, and \Last{} is equipped with a rank/select support.
  We represent $F$ with an array~$C$ of size $\Takai{\sigma}{h-1} \lg n$ bits such that, given a $c \in \Takai{\Sigma}{h-1}$ with its rank~$r_c$,
  $C[r_c]$ is the sum of all symbols in $F$ whose rank is at most the rank of $r_c$.
Each \texttt{\$} in the array~$L$ corresponds to a leaf, and hence to a non-terminal.
Finally, it can be constructed in time linear to the number of nodes~\cite[Thm.~2]{ferragina09xbwt}.
Querying works as follows:
Given a pattern~$P[1..m]$, we proceed like a standard backward search.
It starts with the interval of $P[m]$ in $F$. 
Suppose that we matched $P[i..m]$ with an interval $[b_i..e_i]$.
Then let $f_1 := L.\rank[{P[i-1]}](b_i-1)+1$ and $f_2 := L.\rank[{P[i-1]}](e_i)$.
If $f_1 > f_2$, then there is no path in the trie that reads $P$.
Otherwise, we compute the interval $[C[P[i-1]]+f_1..C[P[i-1]]+f_2]$, and recurse.
We use this operation for finding the interval in $\Takai{\BWT}{1}$ of~$P_p$ by searching $\texttt{\$}P_p$.
The returned range is the range of lexicographic ranks of the non-terminals whose right hand sides have $P_p$ as a prefix.
We conclude that we can find $P_p$ in $|P_p|$ backward search steps on the XBWT\@.
For our running example, where $P_p = \texttt{a}$, we take the interval of all \texttt{A}'s in $F$, 
and then select all \texttt{\$}'s in $F$ within that range. 
The ranks of these \texttt{\$}'s corresponds to the non-terminals \texttt{A}, \texttt{B}, and \texttt{C}.

Finally, we need the colexicographic order of the non-terminals for matching $P_1$ (and building the wavelet tree on the colexicograhically ranked non-terminals of $\Takai{\BWT}{h}$).
For that we have two options:
(a) we create an additional XBWT on the blind tree of the lexicographic sorted right-hand side strings of the non-terminals on height~$h$, or
(b) a simple permutation with $\Takai{\sigma}{h} \lg \Takai{\sigma}{h}$ bits.
The former approach is depicted in \cref{figXBWTLexRank} in the appendix, the latter approach given by \cref{tabColexRank}.

\subsection{Complexity Analysis}
Up so far, we have studied the case that we stop the construction of the grammar at height~$1$.
However, we can build the grammar up to a height~$t_T = \Oh{\lg n}$, and then build $\Takai{\BWT}{t_T}$ on $\Takai{T}{t_T}$.
We then store for each height~$h$ a separate XBWT equipped with the wavelet tree of \citet{barbay14wavelet} supporting a query in \Oh{\lg \lg \Takai{\sigma}{h}} time.
The final BWT can be represented by a data structure supporting partial rank queries~\cite{belazzougui14index} in constant time
such that we can find a core in $\Takai{T}{t_T-1}$ of length~$\ell = \Oh{|P|/2^{t_T}}$ in \Oh{\ell} time.
For the interval in BWT containing the occurrences of $P_p$, there are now not two, but $2^{t_T}$ possibilities:
This is because, for each recursive application of the GCIS grammar, we have the possibility to include the last run of symbols of the last LMS factor.
Note that large values of $t_T$ makes it unfeasible to find short patterns that exhibit cores only at lower heights; this shortcoming is addressed in the next section.

Unfortunately, for a meaningful worst case query time analysis, we need to bound the lengths of the LMS factors of $P$.
We can do so if we enhance the grammar to be run-length compressed, i.e., reducing character runs to single characters with their length information.
Then a run-length compressed LMS substring on height~$h$ has a length of at most $2\Takai{\sigma}{h}$, and therefore, we can find a range of non-terminals containing such a string in $\Takai{\sigma}{h} \lg \lg \Takai{\sigma}{h}$ time.
This gives
$\Oh{\sum_{h=0}^{t_T-1} \Takai{\sigma}{h} \lg \lg \Takai{\sigma}{h}}$ time for finding the $2^{t_T}$ initial backward search intervals, and
$\Oh{|P|}$ time for conducting the backward search on all possible intervals.
Although the worst case time is never better than that of the FM-index built directly on \Takai{\BWT}{0},
it can be improved by leveraging parallel executions.
In fact, conducting the backward search on the $2^{t_T}$ possible intervals is embarrassingly parallel.
Given we have $\rho$ processors, we set $t_T$ to $\Oh{\lg \rho}$.
Then each backward search can be handled by each processor individually in \Oh{|P|/\rho} time.
Finally, we merge the results in a tournament tree in $\Oh{\lg \rho}$ time.

The wavelet tree on $\Takai{\BWT}{t_T}$ uses $n H_k/2^{t_T} + \oh{n H_k} + \Oh{n/2^{t_T}}$ bits with the representation of \citet{belazzougui15wavelet},
and the XBWT on height~$h$ takes
$\Takai{g}{h} \lg (\Takai{\sigma}{h-1}+\Takai{\sigma}{h}) + \Oh{\Takai{g}{h}}$, where $\Takai{g}{h}$ is the size of the concatenation of all right hand side rules of $\Takai{\Sigma}{h}$, for each $h \in [1..t_T]$.
The overall construction time is linear to the text length.

\section{Practical Improvements}\label{secPracticalImprovements}
For practical reasons, we follow the aforementioned examples with respect to that we stop the grammar construction at height~$1$.
That is because we experienced that the grammar at height~$1$ already compresses well, while higher levels introduce much more non-terminals outweighing the compression.
Contrary to that, we additionally introduce a \teigi{chunking parameter} $\lambda \in \Oh{\log_\sigma n}$.
This parameter chops each LMS factor into factors of length $\lambda$ with a possibly smaller last factor such that each non-terminal has a length of at most $\lambda$.
The idea for such small $\lambda$ is that we can interpret the right hand side of each non-terminal as an integer fitting into a constant number of machine words.
For the dictionary on the right hand sides of the non-terminals, we drop the idea of the XBWT, 
but use compressed bit vectors \bv{F} and \bv{R}, each of length $\sigma^\lambda$.
We represent $\pi(X)$ for each non-terminal as an integer~$v \in [1..\sigma^\lambda]$ and store it by setting $\bv{F}[v] = 1$.
Similarly, we represent the reversed string $\pi(X)$ as such an integer~$v'$ and set $\bv{R}[v'] = 1$.
We endow \bv{F} and \bv{R} with rank/select-support data structures.
We additionally store a permutation to convert a value of $\bv{R}.\rank[1]$ to $\bv{F}.\select[1]$.

\subsection{Pattern Matching}
Unfortunately, by limiting the right hand sides of the non-terminals at length~$\lambda$, 
the property that only the first and last non-terminal of the parsed pattern is not in the core no longer holds in general. 
Let again $P = P_1 \cdots P_p$ be the LMS factorization of our pattern.
We assume that $p \ge 2$ and $|P| > \lambda$; the other cases are analyzed afterwards.
For $x \in [2..p]$, we define the chunks $P_{x,1} \cdots P_{x,c_x} = P_x$ with $|P_{x,j}| = \lambda$ for each $j \in [1..c_x-1]$ and $|P_{x,c_x}| \in [1..\lambda]$.
Then, due to the construction of our chunks, there are non-terminals $Y_{x,j} \in \Takai{\Sigma}{1}$ with $\pi(Y_{x,j}) = P_{x,j}$ for all $x \in [2..p-1]$ and $j \in [1..c_x]$.
Hence, $Y_{2,1} \cdots Y_{2,c_2} Y_{3,1} \cdots Y_{p-1}{c_{p-1}}$ is the core of $P$ on height~$1$.
The core can be found as a \BWT{} range analogously as explained in \cref{secPatternMatching}.

But before searching the core, we first find $P_p$. 
We only analyze the case of an occurrence where the last character run in $P_p$ has not been transferred to a new factor.
In that case, we find a range of non-terminals whose right hand sides start with $P_{p,c_p}$.
In detail, we interpret $P_{p,c_p}$ as a binary integer~$v$ having $|P_{p,c_p}|\lg \sigma$ bits. 
Then we create two integers~$v_1,v_2$ by padding $v$ with \bsq{\texttt{0}} and \bsq{\texttt{1}} bits to $v$'s right end (interpreting the right end as the bits encoding the end of the string~$P_{p,c_p}$), respectively, such that $v_1$ and $v_2$ have $\lambda \lg \sigma$ bits with $v_1 \le v_2$.
This gives us the ranks $[\bv{F}.\rank[1](v_1)..\bv{F}.\rank[1](v_2)]$ of all non-terminals whose right hand sides start with $P_{p,c_p}$,
and this interval of ranks translates to a range in \Takai{\BWT}{1}.
Because we know that $P_p$ was always a prefix of a non-terminal in \cref{secPatternMatching},
we can apply the backward search to extend this range to the range of $P_{p,1} \cdots P_{p,c_p}$,
and then continue with searching the core.

Finally, to extend this range to the full pattern,
we remember that an occurrence of $P_1$ in~$T$ was always a suffix of the right hand side of a non-terminal. 
Thus if $P_1 < \lambda$, then we can process analogously.
If not, then such a former right-hand side has been chunked into strings of length~$\lambda$, where the last string has a length in $[1..\lambda]$.
Because we want to match a suffix, we have therefore $\lambda$ different ways in how to chunk $P_1 = P_{1,1} \cdots P_{1,c_1}$ into the same way with $|P_{1,c_1}| \in [1..\lambda]$.
Let us fix one of these chunkings.
We try to extend the range of the core by $P_{1,2} \cdots P_{1,c_1}$ with the backward search steps as before.
If we successfully obtain a range, then we could proceed with $P_{1,1}$ as with $P_1$ in \cref{secPatternMatching} with a top-down traversal of the wavelet tree.
However, here we use the bit vector~$\bv{R}$ and interpret the reverse of $P_{1,1}$ like $P_{p,c_p}$ above as an integer to obtain an interval~$I$ of colexicograhic ranks for all non-terminals whose reversed right-hand sides have the reverse of $P_{1,1}$ as a prefix (i.e., whose right hand sides have $P_{1,1}$ as a suffix). 
Unfortunately, we empirically evaluated that the top-down traversal of the wavelet tree built on the colexicograhically ordered not-terminals is not space-economic in conjunction with the run-length compression of \Takai{\BWT}{1}.
Instead, we have built the wavelet tree with the non-terminals in (standard) lexicographic order, and now use the permutation from \bv{R} to \bv{F} for each element of the interval~$I$, and locate it in the wavelet tree individually.

\subsection{Small Patterns}
Here, we accommodate patterns with $p = 1$ or $|P| < \lambda$,
First, for $p=1$ but $|P| \ge \lambda$, we have $P = P_1$, and we treat $P_1$ exactly like in the above algorithm by trying $\lambda$ different chunkings~$P_1 = P_{1,1} \cdots P_{1,c_1}$
with $|P_{1,c_1}| \in [1..\lambda]$, find all non-terminals having $P_{1,c_1}$ as prefixes of their right hand sides, 
extend the matching interval to a interval of $P_{1,2} \cdots P_{1,c_1}$ via backwards search steps,
and finally use the colexicograhic rankings of \bv{R} to find $P_{1,1} \cdots P_{1,c_1}$.

For $|P| < \lambda$, we need a different data structure:
We create a generalized suffix tree on the right hand sides of all non-terminals.
The \teigi{string label} of a node~$v$ is the concatenation of edge labels read from the root to~$v$.
We augment each node by the number of occurrences of its string label in $T$.
For a given pattern~$P$, we find the highest node~$v$ whose string label has $P$ as a prefix. 
Then the answer to $\fnCount(P)$ is the stored number of occurrences in~$v$.
For the implementation, we represent the generalized suffix tree in LOUDS~\cite{jacobson89rank}, 
and store the occurrences in a plain array by the level order induced by LOUDS\@.

\begin{table*}[th]
  \caption{Comparison of \RLFM{0} and \RLFM{1} on the datasets described in \cref{secExperiments}. 
    The space is in Mebibytes ([MiB]), and \bsq{[M]} denotes mega ($10^6$).
	$\Takai{r}{0}$ and $\Takai{r}{1}$ are the number of character runs in $\Takai{\BWT}{0}$ and $\Takai{\BWT}{1}$, respectively,
	and $\Takai{\sigma}{0}$ and $\Takai{\sigma}{1}$ are, respectively, the number of their different symbols.
      The column $\lg |P|$ is the logarithmic pattern length at which \RLFM{1} starts to become faster than \RLFM{0} on answering $\protect\fnCount{(P)}$.
      See \cref{secPracticalImprovements} for a description of the chunking parameter~$\lambda$.
    }
        \label{tabSpace}
        \centering
        \begin{tabular}{l*{9}{r}}
	  \toprule
	  \multicolumn{3}{c}{input text} & \multicolumn{2}{c}{\RLFM{0}} & \multicolumn{5}{c}{\RLFM{1}} \\
	  \cmidrule(lr){1-3}
	\cmidrule(lr){4-5}
	  \cmidrule(lr){6-10}
\multicolumn{1}{c}{name }&
\multicolumn{1}{c}{space [MiB] }&
\multicolumn{1}{c}{$\sigma$ }&
\multicolumn{1}{c}{$\Takai{r}{0}$ [M] }&
\multicolumn{1}{c}{space [MiB] }&
\multicolumn{1}{c}{$\lambda$ }&
\multicolumn{1}{c}{space [MiB] }&
\multicolumn{1}{c}{$\Takai{\sigma}{1}$ }&
\multicolumn{1}{c}{$\Takai{r}{1}$ [M] }&
\multicolumn{1}{c}{$\lg |P|$ }
\\
            \cline{1-10}
	  \multirow{4}{*}{\textsc{cere}} & \multirow{4}{*}{439.9} & \multirow{4}{*}{6} & \multirow{4}{*}{11.6} & \multirow{4}{*}{26.8} & 1 & 26.5 & 6 & 11.6 & - \\
\cline{6-10}
            & & & & & 4 & 17.3 & 271 & 5.8 & 11 \\
\cline{6-10}
            & & & & & 6 & 15.1 & 1081 & 5.1 & 12 \\
            \cline{6-10}
            & & & & & 7 & 14.9 & 1790 & 5.0 & 13 \\
\hline 
	  \multirow{4}{*}{\textsc{chr19.15}} & \multirow{4}{*}{845.8} & \multirow{4}{*}{6} & \multirow{4}{*}{32.3} & \multirow{4}{*}{70.8} & 1 & 69.7 & 6 & 32.3 & - \\
\cline{6-10}
            & & & & & 4 & 47.1 & 140 & 16.5 & 9 \\
\cline{6-10}
            & & & & & 6 & 40.3 & 620 & 14.3 & 11 \\
            \cline{6-10}
            & & & & & 7 & 39.7 & 1174 & 13.8 & 12 \\
\hline 
	  \multirow{4}{*}{\textsc{e.coli}} & \multirow{4}{*}{107.5} & \multirow{4}{*}{16} & \multirow{4}{*}{15.0} & \multirow{4}{*}{26.2} & 1 & 25.4 & 16 & 15.0 & - \\
\cline{6-10}
            & & & & & 4 & 17.8 & 809 & 7.3 & 13 \\
\cline{6-10}
            & & & & & 6 & 15.3 & 1764 & 6.3 & 13 \\
            \cline{6-10}
            & & & & & 7 & 15.1 & 2356 & 6.2 & 13 \\
\hline
            \multirow{4}{*}{\textsc{para}} & \multirow{4}{*}{409.4} & \multirow{4}{*}{6} & \multirow{4}{*}{15.6} & \multirow{4}{*}{34.4} & 1 & 34.0 & 6 & 15.6 & - \\
\cline{6-10}
            & & & & & 4 & 22.6 & 296 & 7.9 & 11 \\
\cline{6-10}
            & & & & & 6 & 19.6 & 1620 & 6.9 & 12 \\
            \cline{6-10}
            & & & & & 7 & 19.4 & 2701 & 6.7 & 13 \\
\hline
\multirow{3}{*}{\textsc{artificial}.1} & \multirow{3}{*}{502.5} & \multirow{3}{*}{5} & \multirow{3}{*}{50.9} & \multirow{3}{*}{91.4} & 4 & 69.2 & 131 & 28.7 & 8 \\
\cline{6-10}
            & & & & & 6 & 67.4 & 611 & 25.8 & 10 \\
            \cline{6-10}
            & & & & & 7 & 67.1 & 1164 & 25.3 & 11 \\
\hline 
	    \multirow{3}{*}{\textsc{artificial}.2} & \multirow{3}{*}{500.0} & \multirow{3}{*}{5} & \multirow{3}{*}{87.5} & \multirow{3}{*}{141.8} & 4 & 109.9 & 131 & 49.3 & 7 \\
\cline{6-10}
            & & & & & 6 & 107.8 & 611 & 44.3 & 9 \\
            \cline{6-10}
            & & & & & 7 & 107.4 & 1164 & 43.5 & 10 \\
\hline 
            \multirow{3}{*}{\textsc{artificial}.4} & \multirow{3}{*}{495.0} & \multirow{3}{*}{5} & \multirow{3}{*}{147.0} & \multirow{3}{*}{215.8} & 4 & 168.4 & 131 & 81.1 & 7 \\
\cline{6-10}
            & & & & & 6 & 166.0 & 611 & 72.6 & 9 \\
            \cline{6-10}
            & & & & & 7 & 165.4 & 1164 & 71.3 & 10 \\
\hline 
	    \multirow{3}{*}{\textsc{artificial}.8} & \multirow{3}{*}{485.0} & \multirow{3}{*}{5} & \multirow{3}{*}{237.4} & \multirow{3}{*}{300.2} & 4 & 235.1 & 131 & 123.4 & 7 \\
            \cline{6-10}
& & & & & 6 & 228.4 & 611 & 109.2 & 9 \\
            \cline{6-10}
            & & & & & 7 & 226.6 & 1164 & 107.0 & 10 \\
\bottomrule
        \end{tabular}
    \end{table*}

    \begin{figure*}[t]
            \centering
            \includegraphics[width=\textwidth]{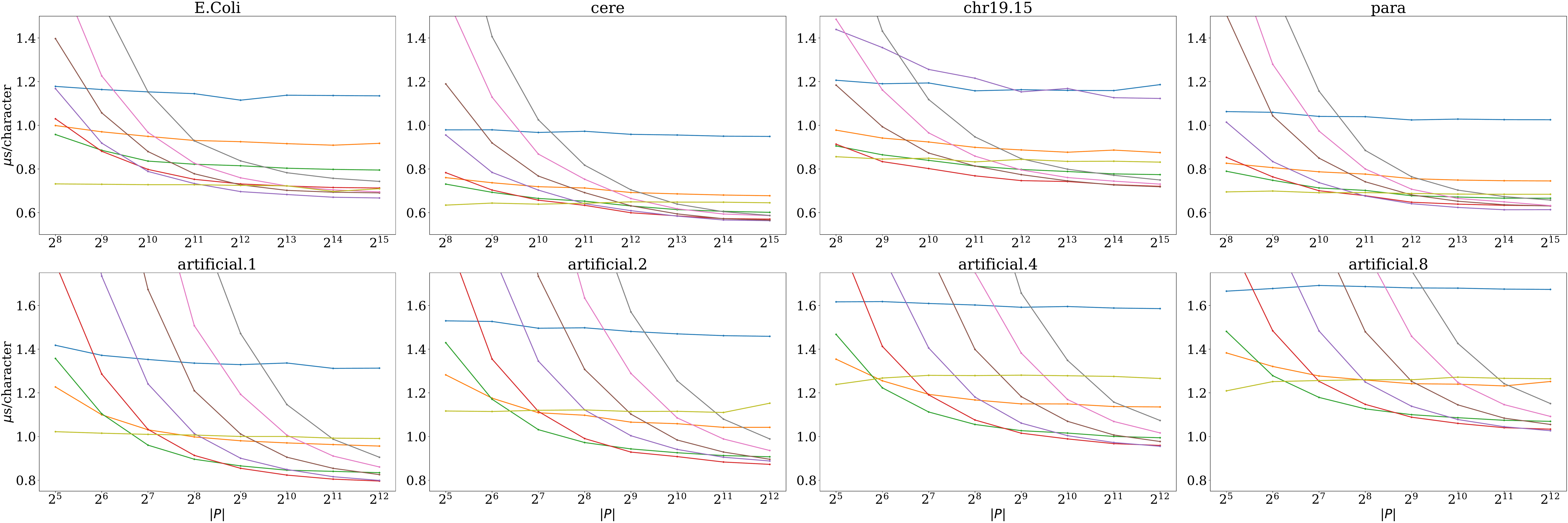}
	    \adjustbox{valign=b}{\begin{minipage}{0.35\linewidth}
	    \caption{Time for answering $\protect\fnCount{(P)}$.}
            \label{figCountingTime}
	    \end{minipage}
	  }\hfill\adjustbox{valign=t}{\begin{minipage}{0.6\linewidth}
	      \includegraphics[width=\linewidth]{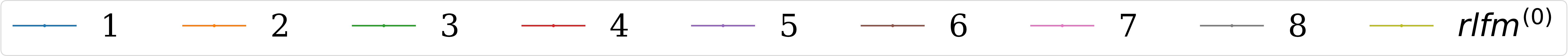}
	    \end{minipage}
	  }
        \end{figure*}

	\section{Implementation and Evaluation}\label{secExperiments}
Our implementation is written in C++17 using the sdsl-lite library~\cite{gog14sdsl}.
The code is available at \url{https://github.com/jamie-jjd/figiss}.

Central to our implementation is the wavelet tree implementation built upon the run-length compressed $\Takai{\BWT}{1}$,
for which we used the class \verb|sdsl::wt_rlmn|.
This class is a wrapper around the actual wavelet tree to make it usable for the RLBWT\@.
Therefore, it is parameterized by a wavelet tree implementation, which we set to \verb|sdsl::wt_ap|, 
an implementation of the alphabet-partitioned wavelet tree of \citet{barbay14wavelet}.
Since we only care about answering \fnCount{}, we do neither sample the suffix array nor its inverse.

The bit vectors $\bv{F}$ and $\bv{R}$ are realized by the class \verb|sdsl::sd_vector<>| leveraging Elias-Fano compression.

\paragraph{Evaluation Environment}
We evaluated all our experiments on a machine with Intel Xeon E3-1231v3 clocked at 3.4GHz
running Ubuntu 20.04.2 LTS\@.
The used compiler was g++ 9.3.0 with compile options \texttt{-std=c++17 -O3}.

\paragraph{Datasets}
We set our focus on DNA sequences, for which we included the datasets
\textsc{cere}, \textsc{Escherichia\_Coli} (abbreviated to \textsc{e.coli}), and \textsc{para} from the repetitive corpus
of Pizza\&Chili\footnote{http://pizzachili.dcc.uchile.cl/repcorpus/real}.
We additionally stored 15 of 1000 sequences of the human chromosome 19\footnote{http://dolomit.cs.tu-dortmund.de/chr19.1000.fa.xz} into the dataset \textsc{chr19.15},
and create a dataset \textsc{artificial}.$x$ for $x \in \{1,2,4,8\}$, 
consisting of a uniform-randomly generated string~$S$ of length $5 \cdot 2^{10}$ on the alphabet $\{\texttt{A},\texttt{C},\texttt{G},\texttt{T}\}$
and 100 copies of $S$, where each character in each copy has been modified by a probability of $x\%$, 
meaning changed to a different character or deleted.
For the experiments we assume that all texts use the byte alphabet. 
In a preprocessing step, after reading an input text~$T$, we reduce the byte alphabet to an alphabet~$\Sigma$
such that each character of~$\Sigma$ appears in $T$. 
We further renumber the characters such that $\Sigma = \{1,\ldots,\sigma\}$  by using a simplified version of \verb!sdsl::byte_alphabet!. 
For technical reasons, we further assume that the texts end with a null byte (at least the used classes in the sdsl need this assumption), which is included in the alphabet sizes~$\sigma$ of our datasets.
We present the characteristics of our datasets in \cref{tabSpace} in the first three columns.

\paragraph{Experiments}
In the following experiments, we call our solution \RLFM{1}, evaluate it for each chunking parameter~$\lambda \in [1..8]$ (cf.~\cref{secPracticalImprovements}),
and compare it with the FM-index \RLFM{0} built on \Takai{\BWT}{0} run-length compressed, 
again without any sampling.\footnote{For space reasons, we only show the evaluation for certain values of $\lambda$. 
The full evaluation is available in the appendix.}
Note that the sampling is only useful for \fnLocate{} queries, and therefore would be only a memory burden in our setting.
While \RLFM{1} uses \verb|sdsl::wt_ap| suitable for larger alphabet sizes, \RLFM{0} uses \verb|sdsl::wt_huff|, a wavelet tree implementation optimized for byte alphabets.
\Cref{tabSpace} shows the space requirements of \RLFM{0} and \RLFM{1}, which are measured by the serialization framework of sdsl.
There, we observe that the larger $\lambda$ gets, the better \RLFM{1} compresses.
However, we are pessimistic that this will be strictly the case for $\lambda > 8$ since the introduced number of symbols exponentially increases while the number of runs~$\Takai{r}{1}$ approaches a saturation curve.
The case $\lambda=1$ can be understood as a baseline: Here, the right-hand sides of all terminals are single characters.
Hence, this approach does not profit from any benefits of our proposed techniques, 
and is provided to measure the overhead of our additional computation (e.g., the dictionary lookups).
Good parameters seem to be $\lambda=4$ and $\lambda=7$, where $\lambda=4$ is faster but uses more space than the solution with $\lambda=7$.
Compared to \RLFM{0}, \RLFM{1} always uses less space, and for the majority of values of $\lambda$, 
answering $\fnCount(P)$ is faster for sufficiently long lengths~$|P|$,
which can be observed in the plots of \cref{figCountingTime}. There, we measure the time for $\fnCount(P)$ with $|P| = 2^{x}$ for each $x \in [8..15]$.
For each data point and each dataset~$T$, we extract $2^{12}$ random samples of equal length from~$T$, perform the query for each sample, and measure the average time per character.\footnote{We extract the patterns from the input such that we can be sure that each pattern actually exists. Non-existing patterns would give \RLFM{1} an advantage since finding the first factor~$P_{1,1}$ take a significant amount of time.}

From \cref{figCountingTime}, we can empirically assess that the larger $\lambda$ is, 
the steeper the falling slope of the average query time per character is for short patterns.
That is because of the split of $P_1$ into $\lambda$ different chunkings.
Our solution with $\lambda=1$ works like \RLFM{0} with some additional overhead and therefore can never be faster than \RLFM{0}.
Interestingly, it seems that the used wavelet tree variant \verb|sdsl::wt_ap| (used for every $\lambda$, in particular for $\lambda=1$) 
seems to be smaller than \verb|sdsl::wt_huff| used for \RLFM{0} regarding the space comparison of \RLFM{1} with $\lambda=1$ and \RLFM{0} in \cref{tabSpace}.
The solution with $\lambda=2$ is only interesting for \textsc{artificial}.$x$, for the other datasets it is always slower than \RLFM{0}.

\section{Future Work}
The chunking into substrings of length~$\lambda$ is rather naive. 
Running a locality sensitive grammar compressor like ESP~\cite{cormode07esp} on the LMS substrings will produce factors of length three
with the property that substrings are factorized in the same way, except maybe at their borders.
Thus, we expect that employing a locality sensitive grammar will reduce the number of symbols and therefore improve $\Takai{r}{1}$.
We further want to parallelize our implementation, and strive to beat \RLFM{0} for smaller pattern lengths.
Also, we would like to conduct our experiments on larger datasets like sequences usually maintained by pangenome indexes of large scale.

\section*{Acknowledgements}
This work was supported by JSPS KAKENHI grant numbers JP21K17701 and JP21H05847.

    \clearpage{}
    \bibliographystyle{abbrvnat}

\clearpage{}
\appendix

\begin{figure}[t]
  \centering{\adjustbox{valign=t}{\tikzstyle{lab} = [midway,fill=white,font=\ttfamily]
\begin{forest}
  for tree={draw,l sep+=1.4em}
[
[,edge label={node[lab]{$\epsilon$}}
[,edge label={node[lab]{a}}
[,edge label={node[lab]{a}}
[,edge label={node[lab]{c}}
[\texttt{A},edge label={node[lab]{\$}}]]]
[,edge label={node[lab]{b}}
[\texttt{B},edge label={node[lab]{\$}}]]
[,edge label={node[lab]{c}}
[\texttt{C},edge label={node[lab]{\$}}]]
]
[,edge label={node[lab]{b}}
[\texttt{D},edge label={node[lab]{\$}}]
[,edge label={node[lab]{c}}
[\texttt{E},edge label={node[lab]{\$}}]]]]
]
\end{forest}
}\hfill\adjustbox{valign=t}{\begin{tabular}{lllll}
	  \toprule
	  F & \Last{} & L  & $\Pi$                     & symbol\\
	  \midrule
	  $\epsilon$ & 0 & \texttt{a } & $\epsilon$ \\
$\epsilon$           & 0 & \texttt{b } & \\
\texttt{a}           & 0 & \texttt{a } & \texttt{a}$\epsilon$ \\
\texttt{a}           & 0 & \texttt{b } & \\
\texttt{a}           & 1 & \texttt{c } & \\
\texttt{a}           & 1 & \texttt{c } & \texttt{aa}$\epsilon$  \\
\texttt{b}           & 0 & \texttt{\$} & \texttt{b}$\epsilon$      &\texttt{D} \\
\texttt{b}           & 1 & \texttt{c } & \\
\texttt{b}           & 1 & \texttt{\$} & \texttt{ba}$\epsilon$     &\texttt{B} \\
\texttt{c}           & 1 & \texttt{\$} & \texttt{ca}$\epsilon$     &\texttt{C} \\
\texttt{c}           & 1 & \texttt{\$} & \texttt{caa}$\epsilon$    &\texttt{A} \\
\texttt{c}           & 1 & \texttt{\$} & \texttt{cb}$\epsilon$     &\texttt{E} \\
\bottomrule
	\end{tabular}
 }
  }\caption{Trie on the right hand sides of all non-terminals of our running example with its XBWT representation, cf.~\cref{figXBWT} for the trie on the reserved right hand sides.
    The ranks of the \texttt{\$} in $L$ corresponds to the colexiographic ranking of the non-terminals, cf.~\cref{tabColexRank}.
  }
  \label{figXBWTLexRank}
\end{figure}

\section{Consistent Grammars}
Our approach is not limited to the GCIS grammar.
We can also make use of a wider range of grammars.
For that purpose, we would like to introduce $\tau$-consistent grammars, and then show how we can use them.
Given an integer $\tau < n$ and a run-length compressed string $T$ of length~$n$,
a set of positions~$S \subset [1..n]$ of $T$ is called \teigi{$\tau$-consistent}
if, for every positions $i,j \in [1..n-\tau+1]$ with $T[i..i+\tau) = T[j..j+\tau)$, it holds that $i \in S$ if and only if $j \in S$
(a $\tau$-synchronizing set~\cite[Def.~3.1]{kempa19synchronizing} is a $2\tau$-consistent set).
A factorizing grammar is \teigi{$\tau$-consistent} if $\Takai{T}{h} = \pi(\Takai{X_{i_1}}{h}) \ldots \pi(\Takai{X_{i_{|\Takai{T}{h+1}|}}}{h})$ 
and the starting positions of the substrings $\pi(\Takai{X_{i_j}}{h})$ for all $j \in [1..|\Takai{T}{h+1}|]$ form a 
$\tau$-consistent set.

Examples of $\tau$-consistent grammars are 
signature encoding~\cite{mehlhorn97signature} with $\tau = \Oh{\lg^* n}$,
the Rsync parse~\cite{gagie19rsync} with a probabilistically selectable $\tau$,
AlgBcp~\cite{ganczorz18edit} with  $\tau = \Oh{1}$,
grammars based on string $\tau$-synchronizing sets~\cite{kempa19synchronizing},
a run-length compressed variant of GCIS with $\tau = 2\sigma'$, 
where $\sigma'$ is the number of different characters in the run-length encoded text.

Now assume that $P$ factorizes into $P = P_1 \cdots P_p$.
If $|P_1|,|P_m| > \tau$, then we can directly apply our approach since $P_2 \cdots P_{m-1}$ can be interpreted as the right-hand sides of non-terminals belonging to the core of~$P$.
Otherwise, let $f$ and $\ell$ be the smallest and largest numbers, respectively such that
$|P_1 \cdots P_f| \ge \tau$ and $|P_\ell \cdots P_p| \ge \tau$.
Then again $P_{f+1} \cdots P_{\ell-1}$ can be found via the core of $P$.
For the other factors, we can proceed analogously as for the chunking into $\lambda$-length substrings described in \cref{secPracticalImprovements}.

\section{Full Experiments}
\begin{figure*}[t]
            \centering
            \includegraphics[width=\textwidth]{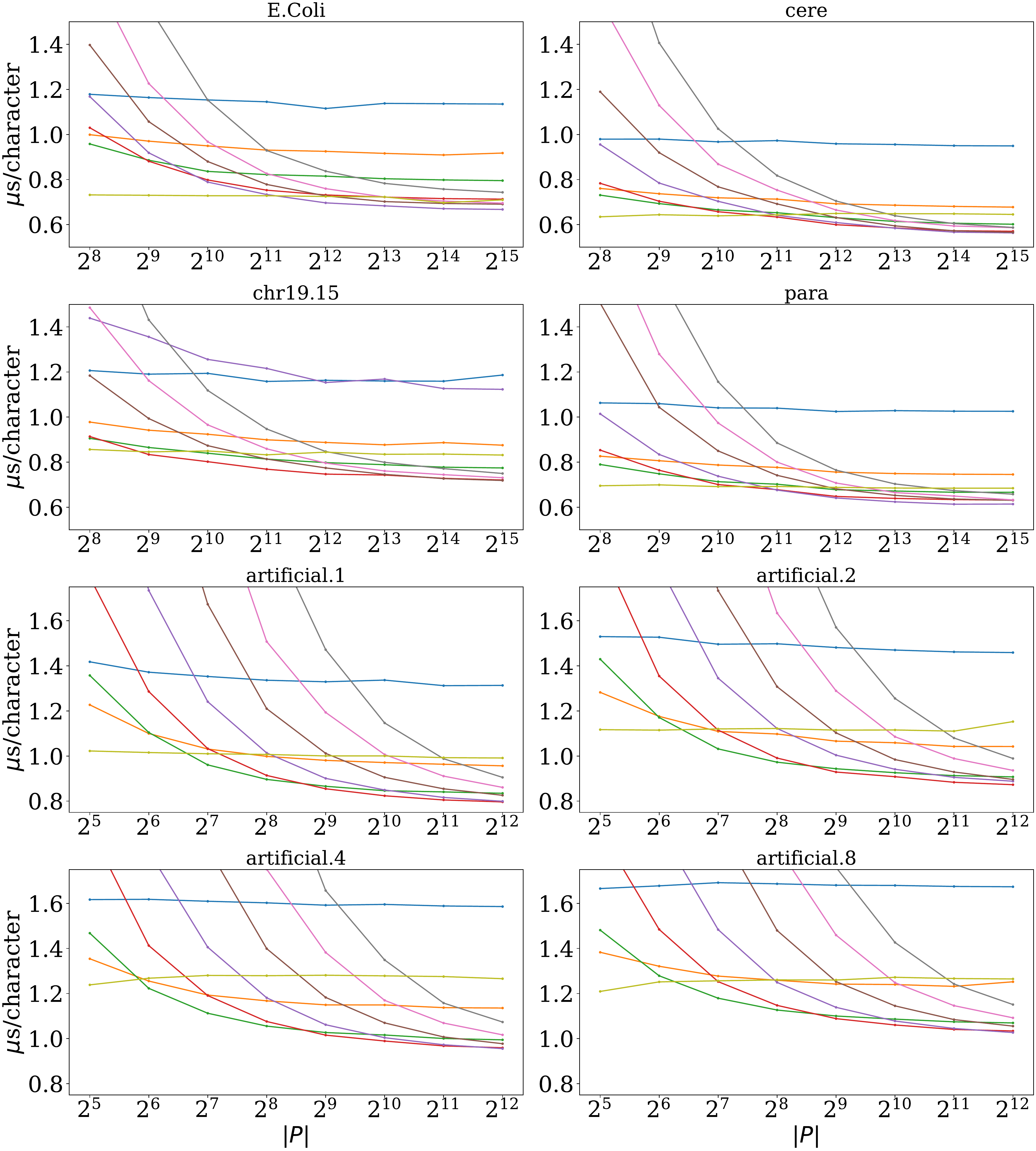}
	      \includegraphics[width=\linewidth]{figure/legend}
	  \caption{Time for answering $\protect\fnCount{(P)}$, cf.~\cref{figCountingTime}}
            \label{figCountingTimeExtended}
        \end{figure*}

Finally, we provide the full experiments (\cref{tabSpacePizzaChili,tabSpaceArtificial})  and plots (\cref{figCountingTimeExtended}) with higher resolution that did not made in into the main text due to space limitations.
We additionally evaluated in \cref{tabConstructionTimeReal,tabConstructionTimeRandom} the construction times for \RLFM{0}, \RLFM{1}, and the FM-index on the plain \Takai{\BWT}{0}.
There, we used the same wavelet tree implementation \verb!sdsl::wt_huff! for the FM-index as for \RLFM{0}.
We observe that the best construction times of \RLFM{1} are roughly 2 -- 3 times slower than for \RLFM{0} and the FM-index.
The construction is the slowest for $\lambda=1$ (up to 10 times slower), and fastest for a $\lambda \in [5..8]$ (the exact number differs for each dataset).

	\begin{table*}[t]
        \caption{Comparison of \RLFM{0} and \RLFM{1} on the real-world datasets described in \cref{secExperiments}.
	See \cref{tabSpace} for a description of the columns.
      }
        \label{tabSpacePizzaChili}
        \centering
        \begin{tabular}{l*{9}{r}}
	  \toprule
	  \multicolumn{3}{c}{input text} & \multicolumn{2}{c}{\RLFM{0}} & \multicolumn{5}{c}{\RLFM{1}} \\
	  \cmidrule(lr){1-3}
	\cmidrule(lr){4-5}
	  \cmidrule(lr){6-10}
\multicolumn{1}{c}{name }&
\multicolumn{1}{c}{space [MiB] }&
\multicolumn{1}{c}{$\sigma$ }&
\multicolumn{1}{c}{$\Takai{r}{0}$ [M] }&
\multicolumn{1}{c}{space [MiB] }&
\multicolumn{1}{c}{$\lambda$ }&
\multicolumn{1}{c}{space [MiB] }&
\multicolumn{1}{c}{$\Takai{\sigma}{1}$ }&
\multicolumn{1}{c}{$\Takai{r}{1}$ [M] }&
\multicolumn{1}{c}{$\lg |P|$ }
\\
            \midrule
            \multirow{8}{*}{\textsc{cere}} & \multirow{8}{*}{439.9} & \multirow{8}{*}{6} & \multirow{8}{*}{11.6} & \multirow{8}{*}{26.8} & 1 & 26.5 & 6 & 11.6 & - \\
            \cline{6-10}
            & & & & & 2 & 20.4 & 26 & 8.3 & - \\
            \cline{6-10}
            & & & & & 3 & 18.4 & 95 & 6.7 & 12 \\
            \cline{6-10}
            & & & & & 4 & 17.3 & 271 & 5.8 & 11 \\
            \cline{6-10}
            & & & & & 5 & 16.7 & 602 & 5.3 & 11 \\
            \cline{6-10}
            & & & & & 6 & 15.1 & 1081 & 5.1 & 12 \\
            \cline{6-10}
            & & & & & 7 & 14.9 & 1790 & 5.0 & 13 \\
            \cline{6-10}
            & & & & & 8 & 14.9 & 2810 & 4.9 & 13 \\
            \midrule 
            \multirow{8}{*}{\textsc{chr19.15}} & \multirow{8}{*}{845.8} & \multirow{8}{*}{6} & \multirow{8}{*}{32.3} & \multirow{8}{*}{70.8} & 1 & 69.7 & 6 & 32.3 & - \\
            \cline{6-10}
            & & & & & 2 & 54.9 & 22 & 23.5 & - \\
            \cline{6-10}
            & & & & & 3 & 50.8 & 58 & 19.0 & 10 \\
            \cline{6-10}
            & & & & & 4 & 47.1 & 140 & 16.5 & 9 \\
            \cline{6-10}
            & & & & & 5 & 44.9 & 305 & 15.1 & - \\
            \cline{6-10}
            & & & & & 6 & 40.3 & 620 & 14.3 & 11 \\
            \cline{6-10}
            & & & & & 7 & 39.7 & 1174 & 13.8 & 12 \\
            \cline{6-10}
            & & & & & 8 & 39.4 & 2086 & 13.6 & 13 \\
            \midrule 
            \multirow{8}{*}{\textsc{E.Coli}} & \multirow{8}{*}{107.5} & \multirow{8}{*}{16} & \multirow{8}{*}{15.0} & \multirow{8}{*}{26.2} & 1 & 25.4 & 16 & 15.0 & - \\
            \cline{6-10}
            & & & & & 2 & 20.4 & 123 & 10.7 & - \\
            \cline{6-10}
            & & & & & 3 & 19.3 & 399 & 8.5 & - \\
            \cline{6-10}
            & & & & & 4 & 17.8 & 809 & 7.3 & 13 \\
            \cline{6-10}
            & & & & & 5 & 17.2 & 1272 & 6.7 & 12 \\
            \cline{6-10}
            & & & & & 6 & 15.3 & 1764 & 6.3 & 13 \\
            \cline{6-10}
            & & & & & 7 & 15.1 & 2356 & 6.2 & 13 \\
            \cline{6-10}
            & & & & & 8 & 15.1 & 3251 & 6.1 & - \\
            \midrule
            \multirow{8}{*}{\textsc{para}} & \multirow{8}{*}{409.4} & \multirow{8}{*}{6} & \multirow{8}{*}{15.6} & \multirow{8}{*}{34.4} & 1 & 34.0 & 6 & 15.6 & - \\
            \cline{6-10}
            & & & & & 2 & 26.5 & 26 & 11.3 & - \\
            \cline{6-10}
            & & & & & 3 & 24.5 & 96 & 9.1 & 12 \\
            \cline{6-10}
            & & & & & 4 & 22.6 & 296 & 7.9 & 11 \\
            \cline{6-10}
            & & & & & 5 & 20.0 & 774 & 7.2 & 11 \\
            \cline{6-10}
            & & & & & 6 & 19.6 & 1620 & 6.9 & 12 \\
            \cline{6-10}
            & & & & & 7 & 19.4 & 2701 & 6.7 & 13 \\
            \cline{6-10}
            & & & & & 8 & 19.3 & 4013 & 6.7 & 14 \\
            \bottomrule
        \end{tabular}
    \end{table*}
    
    \begin{table*}[t]
	\caption{Comparison of \RLFM{0} and \RLFM{1} on the datasets of random generated DNA sequences.
	See \cref{tabSpace} for a description of the columns.}
        \label{tabSpaceArtificial}
        \centering
        \begin{tabular}{l*{9}{r}}
	  \toprule
	  \multicolumn{3}{c}{input text} & \multicolumn{2}{c}{\RLFM{0}} & \multicolumn{5}{c}{\RLFM{1}} \\
	  \cmidrule(lr){1-3}
	\cmidrule(lr){4-5}
	  \cmidrule(lr){6-10}
\multicolumn{1}{c}{name }&
\multicolumn{1}{c}{space [MiB] }&
\multicolumn{1}{c}{$\sigma$ }&
\multicolumn{1}{c}{$\Takai{r}{0}$ [M] }&
\multicolumn{1}{c}{space [MiB] }&
\multicolumn{1}{c}{$\lambda$ }&
\multicolumn{1}{c}{space [MiB] }&
\multicolumn{1}{c}{$\Takai{\sigma}{1}$ }&
\multicolumn{1}{c}{$\Takai{r}{1}$ [M] }&
\multicolumn{1}{c}{$\lg |P|$ }
\\
            \midrule
	    \multirow{8}{*}{\textsc{artificial}.1} & \multirow{8}{*}{502.5} & \multirow{8}{*}{5} & \multirow{8}{*}{50.9} & \multirow{8}{*}{91.4} & 1 & 89.5 & 5 & 50.9 & - \\
            \cline{6-10}
            & & & & & 2 & 84.9 & 17 & 39.0 & 8 \\
            \cline{6-10}
            & & & & & 3 & 71.2 & 51 & 32.3 & 7 \\
            \cline{6-10}
            & & & & & 4 & 69.2 & 131 & 28.7 & 8 \\
            \cline{6-10}
            & & & & & 5 & 68.0 & 297 & 26.7 & 9 \\
            \cline{6-10}
            & & & & & 6 & 67.4 & 611 & 25.8 & 10 \\
            \cline{6-10}
            & & & & & 7 & 67.1 & 1164 & 25.3 & 11 \\
            \cline{6-10}
            & & & & & 8 & 66.9 & 2058 & 25.1 & 11 \\
            \midrule 
	  \multirow{8}{*}{\textsc{artificial}.2} & \multirow{8}{*}{500.0} & \multirow{8}{*}{5} & \multirow{8}{*}{87.5} & \multirow{8}{*}{141.8} & 1 & 138.5 & 5 & 87.5 & - \\
            \cline{6-10}
            & & & & & 2 & 131.5 & 17 & 67.0 & 7 \\
            \cline{6-10}
            & & & & & 3 & 111.8 & 51 & 55.5 & 7 \\
            \cline{6-10}
            & & & & & 4 & 109.9 & 131 & 49.3 & 7 \\
            \cline{6-10}
            & & & & & 5 & 108.5 & 297 & 45.9 & 9 \\
            \cline{6-10}
            & & & & & 6 & 107.8 & 611 & 44.3 & 9 \\
            \cline{6-10}
            & & & & & 7 & 107.4 & 1164 & 43.5 & 10 \\
            \cline{6-10}
            & & & & & 8 & 107.2 & 2069 & 43.2 & 11 \\
            \midrule 
	  \multirow{8}{*}{\textsc{artificial}.4} & \multirow{8}{*}{495.0} & \multirow{8}{*}{5} & \multirow{8}{*}{147.0} & \multirow{8}{*}{215.8} & 1 & 210.1 & 5 & 147.0 & - \\
            \cline{6-10}
            & & & & & 2 & 198.7 & 17 & 111.4 & 6 \\
            \cline{6-10}
            & & & & & 3 & 169.7 & 51 & 91.8 & 6 \\
            \cline{6-10}
            & & & & & 4 & 168.4 & 131 & 81.1 & 7 \\
            \cline{6-10}
            & & & & & 5 & 167.0 & 297 & 75.4 & 8 \\
            \cline{6-10}
            & & & & & 6 & 166.0 & 611 & 72.6 & 9 \\
            \cline{6-10}
            & & & & & 7 & 165.4 & 1164 & 71.3 & 10 \\
            \cline{6-10}
            & & & & & 8 & 165.0 & 2077 & 70.7 & 11 \\
            \midrule 
	\multirow{8}{*}{\textsc{artificial}.8} & \multirow{8}{*}{485.0} & \multirow{8}{*}{5} & \multirow{8}{*}{237.4} & \multirow{8}{*}{300.2} & 1 & 290.9 & 5 & 237.4 & - \\
            \cline{6-10}
            & & & & & 2 & 286.4 & 17 & 174.3 & 8 \\
            \cline{6-10}
            & & & & & 3 & 239.6 & 51 & 141.3 & 7 \\
            \cline{6-10}
            & & & & & 4 & 235.1 & 131 & 123.4 & 7 \\
            \cline{6-10}
            & & & & & 5 & 231.0 & 297 & 113.9 & 8 \\
            \cline{6-10}
            & & & & & 6 & 228.4 & 611 & 109.2 & 9 \\
            \cline{6-10}
            & & & & & 7 & 226.6 & 1164 & 107.0 & 10 \\
            \cline{6-10}
            & & & & & 8 & 225.5 & 2079 & 106.1 & 11 \\
            \bottomrule
        \end{tabular}
    \end{table*}

    \begin{table}[t]
      \caption{Construction times on the real-world datasets. Times are in seconds ([s]).}
        \label{tabConstructionTimeReal}
        \centering
        \begin{tabular}{l*{4}{r}}
	  \toprule
	  \multirow{2}{*}{dataset} & FM-index & \RLFM{0} & \multicolumn{2}{c}{\RLFM{1}} \\
	  \cmidrule(lr){4-5}
	& time [s] & time [s] & $\lambda$ & time [s] \\
        \midrule
	  \multirow{8}{*}{\textsc{cere}} & \multirow{8}{*}{101.6} & \multirow{8}{*}{102.6} & 1 & 881.3 \\
        \cline{4-5}
        & & & 2 & 471.4 \\
        \cline{4-5}
        & & & 3 & 365.0 \\
        \cline{4-5}
        & & & 4 & 299.4 \\
        \cline{4-5}
        & & & 5 & 290.4 \\
        \cline{4-5}
        & & & 6 & 277.4 \\
        \cline{4-5}
        & & & 7 & 275.9 \\
        \cline{4-5}
        & & & 8 & 276.6 \\
        \midrule
	  \multirow{8}{*}{\textsc{chr19.15}} & \multirow{8}{*}{207.7} & \multirow{8}{*}{208.9} & 1 & 2093.7 \\
        \cline{4-5}
        & & & 2 & 1107.7 \\
        \cline{4-5}
        & & & 3 & 807.3 \\
        \cline{4-5}
        & & & 4 & 683.0 \\
        \cline{4-5}
        & & & 5 & 633.2 \\
        \cline{4-5}
        & & & 6 & 626.6 \\
        \cline{4-5}
        & & & 7 & 609.6 \\
        \cline{4-5}
        & & & 8 & 597.2 \\
        \midrule
	  \multirow{8}{*}{\textsc{e.coli}} & \multirow{8}{*}{24.5} & \multirow{8}{*}{25.0} & 1 & 187.1 \\
        \cline{4-5}
        & & & 2 & 110.3 \\
        \cline{4-5}
        & & & 3 & 87.1 \\
        \cline{4-5}
        & & & 4 & 71.9 \\
        \cline{4-5}
        & & & 5 & 66.7 \\
        \cline{4-5}
        & & & 6 & 65.8 \\
        \cline{4-5}
        & & & 7 & 66.3 \\
        \cline{4-5}
        & & & 8 & 71.1 \\
        \midrule
	  \multirow{8}{*}{\textsc{para}} & \multirow{8}{*}{98.0} & \multirow{8}{*}{98.0} & 1 & 755.7 \\
        \cline{4-5}
        & & & 2 & 410.5 \\
        \cline{4-5}
        & & & 3 & 319.3 \\
        \cline{4-5}
        & & & 4 & 267.4 \\
        \cline{4-5}
        & & & 5 & 260.9 \\
        \cline{4-5}
        & & & 6 & 251.5 \\
        \cline{4-5}
        & & & 7 & 248.2 \\
        \cline{4-5}
        & & & 8 & 252.2 \\
        \bottomrule
        \end{tabular}
        \end{table}

	\begin{table}[t]
	  \caption{Construction times for the datasets \textsc{artificial}.$x$.}
        \label{tabConstructionTimeRandom}
        \centering
        \begin{tabular}{l*{5}{r}}
	  \toprule
	  \multirow{2}{*}{dataset} & FM-index & \RLFM{0} & \multicolumn{2}{c}{\RLFM{1}} \\
	  \cmidrule(lr){4-5}
	& time [s] & time [s] & $\lambda$ & time [s] \\
        \midrule
        \multirow{8}{*}{\textsc{artificial}.1} & \multirow{8}{*}{130.0} & \multirow{8}{*}{131.4} & 1 & 616.1 \\
        \cline{4-5}
        & & & 2 & 339.4 \\
        \cline{4-5}
        & & & 3 & 263.2 \\
        \cline{4-5}
        & & & 4 & 227.4 \\
        \cline{4-5}
        & & & 5 & 224.1 \\
        \cline{4-5}
        & & & 6 & 230.7 \\
        \cline{4-5}
        & & & 7 & 230.5 \\
        \cline{4-5}
        & & & 8 & 233.7 \\
        \midrule
        \multirow{8}{*}{\textsc{artificial}.2} & \multirow{8}{*}{129.2} & \multirow{8}{*}{132.9} & 1 & 578.2 \\
        \cline{4-5}
        & & & 2 & 328.3 \\
        \cline{4-5}
        & & & 3 & 254.5 \\
        \cline{4-5}
        & & & 4 & 218.7 \\
        \cline{4-5}
        & & & 5 & 213.2 \\
        \cline{4-5}
        & & & 6 & 220.5 \\
        \cline{4-5}
        & & & 7 & 220.3 \\
        \cline{4-5}
        & & & 8 & 224.2 \\
        \midrule
        \multirow{8}{*}{\textsc{artificial}.4} & \multirow{8}{*}{130.5} & \multirow{8}{*}{137.1} & 1 &  555.2 \\
        \cline{4-5}
        & & & 2 & 555.2 \\
        \cline{4-5}
        & & & 3 & 243.4 \\
        \cline{4-5}
        & & & 4 & 213.6 \\
        \cline{4-5}
        & & & 5 & 209.7 \\
        \cline{4-5}
        & & & 6 & 218.1 \\
        \cline{4-5}
        & & & 7 & 216.5 \\
        \cline{4-5}
        & & & 8 & 222.8 \\
        \midrule
        \multirow{8}{*}{\textsc{artificial}.8} & \multirow{8}{*}{132.4} & \multirow{8}{*}{144.4} & 1 & 530.6 \\
        \cline{4-5}
        & & & 2 & 306.4 \\
        \cline{4-5}
        & & & 3 & 241.0 \\
        \cline{4-5}
        & & & 4 & 211.0 \\
        \cline{4-5}
        & & & 5 & 207.1 \\
        \cline{4-5}
        & & & 6 & 214.9 \\
        \cline{4-5}
        & & & 7 & 215.5 \\
        \cline{4-5}
        & & & 8 & 219.7 \\
        \midrule
        \end{tabular}
    \end{table}

\end{document}